\documentclass[12pt]{article}
\def\bSig\mathbf{\Sigma}

\usepackage[figuresright]{rotating}
\usepackage{amsmath}
\usepackage{comment,bm}
\usepackage{amssymb,latexsym,graphicx}
\usepackage{natbib}
\usepackage{setspace}
\usepackage{adjustbox}
\usepackage[hidelinks]{hyperref}
\usepackage{xcolor,multirow}
\usepackage[ruled,vlined]{algorithm2e}
    \usepackage{amsfonts}
    \usepackage{stmaryrd}
    \usepackage{url}

\def\x{\mathbf{x}}

\def\W{\mathbf{W}}

\def\U{\mathbf{U}}
\def\u{\mathbf{u}}
\def\X{\mathbf{X}}
\def\XX{\mathbb{X}}

\def\b{\mathbf{b}}

\def\BB{\mathbb{B}}
\def\r{\mathbf{r}}

\begin{document}
\def\spacingset#1{\renewcommand{\baselinestretch}%
{#1}\small\normalsize} \spacingset{1}




  \title{\bf Multiway sparse distance weighted discrimination}
  \author{Bin Guo$^1$, Lynn E. Eberly$^{1,2}$, Pierre-Gilles Henry$^2$,\\ Christophe Lenglet$^2$, Eric F. Lock$^1$\\ \\
	   $^1$ Division of Biostatistics, School of Public Health \\
	   $^2$ Center for Magnetic Resonance Research \\
	   University of Minnesota}
	   \date{}
  \maketitle

\begin{abstract}

Modern data often take the  form  of  a  multiway  array.   However, most classification methods are designed for vectors, i.e., 1-way arrays.  Distance weighted discrimination (DWD) is a popular high-dimensional classification method that has been extended to the multiway context,  with  dramatic  improvements  in  performance  when  data  have  multiway  structure.  However,  the  previous implementation of multiway DWD was restricted to classification of matrices, and did not  account  for  sparsity.  In  this  paper,  we  develop  a  general  framework  for  multiway  classification  which  is  applicable  to  any  number  of dimensions and any degree of sparsity. We conducted extensive simulation studies, showing that our model is robust to the degree of sparsity and improves classification accuracy when the data have multiway structure.  For our motivating application, magnetic resonance spectroscopy (MRS) was used to measure the abundance of several metabolites  across  multiple  neurological  regions  and  across  multiple  time  points  in a mouse model  of  Friedreich’s ataxia,  yielding  a  four-way  data  array.  Our method reveals a robust and interpretable multi-region metabolomic signal that discriminates the groups of interest. We also successfully apply our method to gene expression time course data for multiple sclerosis treatment.  An R implementation is available in the package {\tt MultiwayClassification} at \url{http://github.com/lockEF/MultiwayClassification}.

\end{abstract}


\noindent%
\vfill
 
\maketitle

\spacingset{1.45} 
\section{Introduction}
\label{s:intro}

Development and wide deployment of advanced technologies have produced tools that generate massive amounts of data with complex structures. These data are often represented as a multiway or multidimensional array, which extends the two-way data matrix to higher dimensions. This paper concerns the task of classification from multiway data. 
As our motivating data application we consider magnetic resonance spectroscopy (MRS) data for a study of Friedreich's Ataxia in mice. The MRS data measures the concentration of several metabolites across multiple regions of the brain, and across multiple time points after a treatment, yielding a four-way data array: mice $\times$ metabolites $\times$ regions $\times$ time.  We are interested in identifying signal that distinguishes the treatment groups using the totality of the MRS data.   

 A naive approach is to transform the multi-dimensional array to a vector, and then apply high-dimensional classifiers designed for vector-valued data to the transformed vector. However, the effects of the same metabolite in different brain regions and at different time points on the classification into the two classes are very likely correlated. Ignoring the dependence across different dimensions may result in inaccurate classifications and complicate interpretation. Thus, the multiway structure should be considered in the model.  In addition, only some metabolites may be useful for distinguishing the classes, and these distinctions may only be present for some time points.  Exclusion of uninformative features can improve classification performance and interpretation, and this motivates an approach that also accommodates a sparse structure.  In what follows we briefly review existing high-dimensional classifiers (Section \ref{ss:high-dim}), sparse high-dimensional classifiers (Section \ref{ss:sparse}), and classifiers for multiway data (Section \ref{ss:multiway}); our methodological contributions are summarized in Section \ref{ss:contributions}.

\subsection{High-dimensional classification of vectors}
\label{ss:high-dim}
Traditional approaches to classification like logistic regression and Fisher's linear discriminant analysis (LDA) are prone to overfitting with a large number of features, and this has motivated several classification methods for high-dimensional vector-valued data. We can roughly divide these methods into two categories: non-linear classifiers such as k-nearest neighbor classification \citep{cover1967nearest} and random forests \citep{breiman2001random}, and linear classifiers such as penalized  LDA \citep{witten2011penalized}, support vector machines (SVM) \citep{cortes1995support}, and distance weighted discrimination (DWD) \citep{marron2007distance}. Non-linear classifiers are flexible to use as they require minimal assumptions, but may not have straightforward interpretations. In contrast, linear classifiers use a weighted sum of the measured features. SVM and DWD are both commonly used in biomedical research. However, DWD tends to outperform SVM in scenarios with high-dimensional data and a lower sample size;  here SVM suffers from the data piling problem, which means many cases will pile up at the discriminating margins as a symptom of overfitting \citep{marron2007distance}. 

\subsection{High-dimensional classification of vectors with sparsity}
\label{ss:sparse}
 
 In their original formulations, SVM, DWD, and other linear classification approaches use all available variables. However, in practice there may be only a few important variables affecting the outcome,  especially in biomedical applications to disease classification with imaging or genomics data \citep{zou2019classification}.  Thus, methods that use all variables for classification may include too many noise features and deteriorate the classification performance due to error accumulation \citep{fan2008high}.  Sparse methods, in which only a subset of variables are used for classification, can improve performance in this respect and also improve interpretation by identifying a small number of informative features. 
\cite{zou2019classification} gave a detailed review of high-dimensional classification methods that can account for sparsity. Many sparse classifiers were derived based on high dimensional extensions of the classical discriminant analysis, such as linear programming discriminant (LPD) \citep{cai2011direct}, penalized LDA \citep{witten2011penalized}, regularized optimal affine discriminant (ROAD) \citep{fan2012road} and direct sparse discriminant analysis (DSDA) \citep{mai2012direct}. Other methods have extended margin-based approaches like SVM and DWD. \cite{hastie2009elements} gives a nice complete introduction of SVM from its geometric view to its statistical loss $+$ penalty formulation. A class of methods extended SVM to enforce sparsity by using different loss functions or penalties, such as lasso penalized SVM \citep{bradley1998feature}, elastic-net penalized SVM \citep{wang2008hybrid} and SCAD penalized SVM \citep{zhang2006gene}.  For DWD,  the objective function can also be represented as an analogous loss $+$ penalty formulation \citep{liu2011hard}. \cite{wang2016sparse} proposed sparse DWD (SDWD) by adding lasso and elastic-net penalties on the model coefficients, which can improve performance and efficiency of DWD in high-dimensional classification.

\subsection{High-dimensional classification of multiway arrays} 
\label{ss:multiway}

There is a growing literature on multiway classification by extending classifiers of vectors to multiway arrays using factorization and dimension reduction techniques. 
\cite{ye2004two} and \cite{bauckhage2007robust} extended LDA and related approaches to multiway data. \cite{tao2005supervised} proposed a supervised tensor learning framework by performing a rank-1 decomposition on the coefficients to reduce dimension, in which the coefficients are factorized into a single set of weights for each dimension. \cite{wimalawarne2016theoretical} investigated tensor-based classification where a logistic loss function and a penalty term with different continuous tensor norms for the coefficients are considered. \cite{pan2018covariate} developed a classification approach that allows for multiway data and covariate adjustment; their proposal, termed Covariate-Adjusted Tensor Classification in High Dimensions (CATCH), assumes a multiway structure in the residual covariance but not in the signals discriminating the classes.     \cite{lyu2017discriminating} proposed multiway versions of DWD and SVM under the assumption that the coefficient array is low-rank. Their implementation of multiway DWD was shown to dramatically improve performance over two-way classifiers when the data have multiway structure, and also tended to outperform analogous extensions of SVM. However, their method is restricted to use for three-way data and does not account for sparsity.

\subsection{High-dimensional classification of multiway arrays with sparsity: present contributions}
\label{ss:contributions}
In this paper, we developed a general framework for multiway classification that is applicable to any number of dimensions and any degree of sparsity. Our proposed approach builds a connection between the two approaches of multiway DWD and sparse DWD. The central assumption is that the signal discriminating the groups can be efficiently represented by meaningful patterns in each dimension, which we identify by imposing a low-rank structure on the coefficient array.
Adding penalty terms to multiway DWD can enforce sparsity and give better performance even when the data are not sparse. The rest of the paper is organized as follows. In Section~\ref{sec:framework} we introduce our formal mathematical framework and notation.  In Section~\ref{sec:existing} we briefly review the standard DWD, sparse DWD and multiway DWD  methods. In Section~\ref{sec:proposed} we describe our proposed multiway sparse DWD method and algorithms for estimation. We show the multiway approach has improved performance and interpretation compared with existing methods through simulation studies (Section~\ref{sec:simulations}). We considered two data applications in Section~\ref{sec:application}, and these results showed that the multiway sparsity model is robust to data with any degree of sparsity and has competitive classification performance over other classification techniques. The article concludes with discussion and future directions in Section~\ref{sec:discussion}.

\section{Notation and framework}
\label{sec:framework}

Throughout this article bold lowercase characters ($\mathbf{a}$) denote vectors, bold uppercase characters ($\mathbf{A}$) denote matrices, and blackboard bold uppercase characters ($\mathbb{A}$) denote multiway arrays of the specified dimension (e.g., $\mathbb{A}: P_1 \times P_2 \times \cdots \times P_K$).  Square brackets index entries within an array, e.g., $\mathbb{A}[p_1, p_2,\cdots,p_K]$.  The operators $||\cdot||_1$ and $||\cdot||_2$ define the generalization of the $L_1$ and $L_2$ norms, respectively: 
\[||\mathbb{A}||_1 = \sum_{p_1=1}^{P_1} \cdots \sum_{p_K=1}^{P_K} |\mathbb{A}[p_1, p_2,\cdots,p_K]| \; \; , \; \; ||\mathbb{A}||_2^2 = \sum_{p_1=1}^{P_1} \cdots \sum_{p_K=1}^{P_K} \mathbb{A}[p_1, p_2,\cdots,p_K]^2.   \]
The generalized inner product for two arrays $\mathbb{A}$ and $\mathbb{B}$  of the same dimension is 
\[\mathbb{A} \cdot \mathbb{B} = \sum_{p_1=1}^{P_1} \cdots \sum_{p_K=1}^{P_K} \mathbb{A}[p_1,\hdots,p_K] \mathbb{B}[p_1,\hdots,p_K],\]
the generalized outer product for $\mathbb{A}: P_1\times \cdots \times P_K$ and $\mathbb{B}: Q_1 \times \cdots \times Q_L$ is $\mathbb{A} \circ \mathbb{B}: P_1 \times \cdots \times P_K \times Q_1 \dots Q_L$ where
\[\mathbb{A} \circ \mathbb{B}[p_1,\hdots,p_K,q_1,\hdots,q_L]=\mathbb{A}[p_1,\hdots,p_K]\mathbb{B}[q_1,\hdots,q_L].\]

For our context, $\mathbb{X}$: $N \times P_1 \times \cdots \times P_K$ gives data in the form of a $K$-way array for $N$ subjects, where $P_k$ is the size of the $k^{th}$ dimension for $k=1,\hdots,K$.  Each subject belongs to one of two classes denoted by $-1$ and $+1$; let $y_i \in \{-1,1\}$ give the class label for each subject and $\mathbf{y}=[y_1, \hdots, y_N ]$.  Our goal is to predict the class labels $\mathbf{y}$ based on the multiway covariates $\mathbb{X}$. 

\section{DWD and its extensions}
\label{sec:existing}

\subsection{Distance Weighted Discrimination (DWD)}
\label{standardDWD}

Here we briefly describe the standard DWD for high-dimensional vector-valued data $\x_i$ for each subject, given by the rows of $\mathbf{X}:N \times P$. 
The goal of DWD and related methods is to find the hyperplane $\mathbf{b} = [b_1, \cdots , b_P]$ which best separates the two classes via the subject \emph{scores} $\X \b$. To solve this standard binary classification problem, 
SVM \citep{cortes1995support} identifies the hyperplane that maximizes the  margin separating the two classes.
However, SVM can suffer from the data piling problem as shown in \cite{marron2007distance}, which means many data points may pile up on the margin when the SVM is applied to high-dimensional data. To tackle this issue, they proposed distance weighted discrimination (DWD) which finds the separating hyperplane that minimizes the sum of the inverse distance from the data points to the hyperplane. The standard DWD is formulated as the following optimization problem:
$$
\underset{\r, \b, b_0, \boldsymbol{\xi}}{\operatorname{argmin}} \sum_{i=1}^N \frac{1}{r_{i}}+C  \xi_i
$$
subject to $r_i=y_i (\x_i^T \b + b_0) + \xi_i \geq 0$ and $\xi_i \geq 0$ for $i=1,\hdots,N$, and $||\b|| \leq 1$.
Here, $C$ is the penalty parameter and $b_0$ is an intercept term.  The classification rule is given by the sign of $\x_i \b + b_0$, and thus $\xi_i$ can be considered a penalty for misclassification.

\subsection{Sparse DWD}
\label{sparseDWD}

The standard DWD may not be suitable for high dimensional classification when the underlying signal is sparse, as it does not conduct variable selection. To further improve performance and efficiency of DWD in high-dimensional classification, \cite{wang2016sparse} proposed sparse DWD by adding penalties on the model coefficients to enforce sparsity. The generalization of standard DWD to sparse DWD is based on the fact that 
the objective function for standard DWD can be decomposed into two components: loss function and penalty \citep{liu2011hard}:
\begin{equation}
\underset{b_0, \b}{\arg \min } \frac{1}{N} \sum_{i=1}^{N} V\left(y_{i}\left(b_0+\mathbf{x}_{i}^{T} \b\right)\right)+\frac{\lambda_{2}}{2}\|\b\|_{2}^{2}    
\label{eq_dwd1}
\end{equation}
where the loss function is given by
\begin{equation}
V(u)=\left\{\begin{array}{ll}
{1-u,} & {\text { if } u \leq 1 / 2} \\
{1 /(4 u),} & {\text { if } u>1 / 2.}
\end{array}\right\}  
\label{eq_dwd2}
\end{equation}
To account for sparsity, many variations of penalties can be added in the model, such as lasso and elastic-net \citep{wang2016sparse}. The elastic-net penalty often outperforms the lasso in prediction, and thus the elastic-net penalized DWD is attractive, with the objective function 
\begin{equation}
\underset{b_0, \b}{\arg \min } \frac{1}{N} \sum_{i=1}^{N} V\left(y_{i}\left(b_0+\mathbf{x}_{i}^{T} \b\right)\right)+P_{\lambda_{1}, \lambda_{2}}(\b)   
\label{eq_dwd3}
\end{equation}
with $V(\cdot)$ defined as in \eqref{eq_dwd2} and 
\begin{equation*}
P_{\lambda_{1}, \lambda_{2}}(\b)=
\lambda_1||\b||_1+\frac{\lambda_{2}}{2}||\b||_2^2
\end{equation*}
in which $\lambda_1$ and $\lambda_2$ are tuning parameters for regularization.  Both parameters control the shrinkage of the coefficients toward $0$.  However, the $L_1$ penalty controlled by $\lambda_1$ may result in some model coefficients being shrunk exactly to 0, and in this way, sparsity (retention of relatively few of the covariates for the classification) is imposed.  In practice, $\lambda_1$ and $\lambda_2$ can be determined by cross-validation.  Comparing the objectives~\eqref{eq_dwd1} and~\eqref{eq_dwd3}, note that the penalized elastic net DWD is equivalent to standard DWD when $\lambda_1=0$.

\subsection{Multiway DWD}
\label{multiwayDWD}

The standard DWD (Section~\ref{standardDWD}) and sparse DWD (Section~\ref{sparseDWD}) are designed for high-dimensional classification on vector-valued data.  \cite{lyu2017discriminating} proposed a multiway DWD model which extends the standard DWD from single vector to multiway features. The assumption of this model is that the multiway coefficient matrix can be decomposed into patterns that are particular to each dimension, giving a low-rank-representation. 

Consider classification of subjects with matrix valued covariates $\X_i: P_1 \times P_2$, concatenated to form the three-way array $\XX: N \times P_1 \times P_2$ (for example, mice by metabolites by brain regions). In this multiway context, the coefficients take the form of a matrix $\mathbf{B}: P_1\times P_2$, and the separating hyperplane is given by 
$
f\left(\mathbf{X}_{i}\right) = \mathbf{X}_{i} \cdot \mathbf{B}
$
where $\mathbf{B}$ is the coefficient matrix. The rank-1 multiway DWD model assumes that the coefficient matrix $\mathbf{B}$ has the rank-1 decomposition: $\mathbf{B}=\mathbf{u}_1 \mathbf{u}_2^\intercal$,
where $\mathbf{u}_1: P_1 \times 1$ and $\mathbf{u}_2: P_2 \times 1$ denote vectors of weights for each dimension.
Under this assumption, the hyperplane to separate the two classes is:
$$
\begin{aligned}
f\left(\mathbf{X}_{i}\right)= \mathbf{X}_{i} \cdot \mathbf{B}=\sum_{i_1=1}^{P_1} \sum_{i_2=1}^{P_2}\mathbf{X}_{i}[i_1,i_2]\mathbf{u}_1[i_1]\mathbf{u}_2[i_2].
\end{aligned}
$$
For example, for MRS data of the form $\X_i: \text{Metabolites} \times \text{Regions}$, $\u_1$ gives a discriminating profile across the metabolites and $\u_2$ weights that profile across the different regions.  The rank $R$ model allows for additional patterns in each dimension via $\mathbf{B}=\mathbf{U}_1 \mathbf{U}_2^\intercal$ where $\U_1: P_1 \times R$ and $\U_2: P_2 \times R$.  
The coefficients are estimated by iteratively updating the weights in each dimension to optimize an objective function. It has been shown that this multiway DWD model can improve classification accuracy when the underlying true model has a multiway structure and can provide a simple and straightforward interpretation \citep{lyu2017discriminating}.

\section{Proposed methods}
\label{sec:proposed}

We propose a general framework for classifying high-dimensional multiway data that combines aspects of sparse DWD (Section~\ref{sparseDWD}) and multiway DWD (M-DWD) (Section~\ref{multiwayDWD}). The proposed method can be considered as a multiway version of sparse DWD that allows for any number of dimensions $\XX_i:  P_1 \times \cdots \times P_K$, combined across subjects to form $\XX: N \times P_1 \times \cdots \times P_K$. In the following we first describe our generalization of multiway DWD to an arbitrary number of dimensions (Section~\ref{ss:genMultDWD}), then we describe our sparsity inducing objective function classification method that assumes the coefficient array has a low-rank decomposition (Section~\ref{ss:objFun}). 

\subsection{Generalized multiway DWD}
\label{ss:genMultDWD}

The generalized rank-1 multiway model assumes that the coefficient array $\mathbb{B}: P_1 \times \cdots \times P_K$ has a rank-1 decomposition $\mathbb{B}=\mathbf{u}_1 \circ \mathbf{u}_2 \circ \cdots \circ  \mathbf{u}_K$, 
where $\{\mathbf{u}_k\}_{k=1}^K$ denote the vector of weights for each dimension. The hyperplane for separating the two classes is $f(\mathbb{X}_i)=\XX_i \cdot \BB$, and thus for the rank-$1$ model 
$$
f(\mathbb{X}_i) = \sum_{i_1=1}^{P_1}\cdots \sum_{i_K=1}^{P_K} \mathbb{X}[i,i_1,\cdots,i_K] \u_1[i_1]\u_2[i_2] \cdots \u_K[i_K].
$$
For our motivating application, $\u_1$ identifies a metabolite profile that is weighted over the regions ($\u_2$) and time points $(\u_3)$ (or equivalently, a time profile that is weighted across the metabolites and regions) to discriminate the classes.

The rank-1 model assumes the classification is given by combining a single pattern in each dimension. However, in practice multiple patterns may contribute, e.g., a different metabolite profile may affect the classification for different regions of the brain. Thus, we propose a rank-R model for the coefficient array $\mathbb{B}$ that is assumed to have a rank-R Candecomp/Parafac (CP) factorization \citep{harshman1970foundations}:  
$$
\mathbb{B}=\llbracket \mathbf{U}_1,\dots,\mathbf{U}_K \rrbracket=\sum_{r=1}^R \mathbf{u}_{1r} \circ \cdots \circ \mathbf{u}_{Kr}
$$
where $\mathbf{U}_k:P_k\times R$ for $k=1,\dots,K$ with columns $\mathbf{u}_{kr}$ as the weight for the $k^{th}$ dimension and $r^{th}$ rank component. The coefficient array $\mathbb{B}$ in the rank-1 multiway model is a special case of the rank-R multiway model when $R = 1$. Moreover, as the CP factorization extends the matrix rank, this model is equivalent to that for the multiway DWD approach in Section~\ref{multiwayDWD} when $K=2$.  Note that with no constraints on the coefficient array $\BB$ the number of free parameters is $\prod_{k=1}^K P_k$, and with the rank-$R$ constraint $\BB$ has $R(P_1+\cdots+P_K)$ free parameters.  Thus, in addition to facilitating interpretation of relevant patterns in each dimension, a low-rank approach can reduce over-fitting and improve performance when there is multiway structure.

\subsection{Objective functions}
\label{ss:objFun}

To allow for sparsity in the generalized multiway DWD model, we consider an extension of the sparse DWD
objective~\eqref{eq_dwd3},  
\begin{equation}
h(\mathbf{y}, \mathbb{X}; \mathbb{B}, b_0)=  \frac{1}{N} \sum_{i=1}^{N} V \left(\mathbf{y}_{i}(b_0+\mathbb{X}_i \cdot \mathbb{B})\right)+P_{\lambda_{1}, \lambda_{2}}(\mathbb{B}),    
\label{eq_genmultdwd}
\end{equation}
which we minimize under the restriction that rank$(\BB)$=R.  For the penalty $P_{\lambda_{1}, \lambda_{2}}(\mathbb{B})$ we consider different extensions of the elastic net to low-rank multiway coefficients, which do or do not distribute the $L_1$ and $L_2$ penalties across the factorization components:
\begin{align*}
    P^\BB_{\lambda_{1}, \lambda_{2}}(\mathbb{B})&=\lambda_1\| \mathbb{B} \|_1 +\frac{\lambda_2}{2} \| \mathbb{B} \|_2^2 \\
    P^\U_{\lambda_{1}, \lambda_{2}}(\mathbb{B})&=\lambda_1\sum_{r=1}^R \prod_{k=1}^K \| \mathbf{u}_{kr}\|_1 +\frac{\lambda_2}{2} \sum_{r=1}^R\prod_{k=1}^K\| \mathbf{u}_{kr} \|_2^2, \text{ or}  \\
     P_{\lambda_{1}, \lambda_{2}}(\mathbb{B})&=\lambda_1\sum_{r=1}^R \prod_{k=1}^K \| \mathbf{u}_{kr}\|_1 +\frac{\lambda_2}{2} \| \mathbb{B} \|_2^2.
\end{align*}
For $R=1$, the three penalties are equivalent.  For $R>1$, we find that the solution of the objective for the penalty $P^\U_{\lambda_{1}, \lambda_{2}}(\mathbb{B})$ is often shrunk to a lower rank than the specified rank $R$ (see Appendix~\ref{sep_l2_sims}). This is due to the distributed $L_2$ penalty term, and similar behavior has been observed in other contexts; for example, when $K=2$ the distributed $L_2$ penalty is equivalent to the nuclear norm penalty for matrices \citep{lock2018tensor}, which favors a smaller rank.  However, the distributed version of the $L_1$ penalty in $P^\U_{\lambda_{1}, \lambda_{2}} (\cdot)$ and $P_{\lambda_{1}, \lambda_{2}}(\cdot)$ is intuitive because it enforces sparsity in the weights for each dimension (e.g., separately for the metabolites, regions, and time points) and is amenable to iterative approaches to optimization (see Section~\ref{optimization}).  Thus, in what follows, we use $P_{\lambda_{1}, \lambda_{2}}(\cdot)$ as our penalty.    

This objective function for generalized sparse multiway DWD subsumes standard DWD, multiway DWD, and sparse DWD. The objective function is equivalent to that for standard DWD when $K=1$ and $\lambda_1=0$ \citep{marron2007distance, liu2011hard}, it is equivalent to sparse DWD \citep{wang2016sparse} when $K=1$ and $\lambda_1>0$, and it is equivalent to multiway DWD \citep{lyu2017discriminating} if $\lambda_1=0$ and $K=2$. 

\subsection{Optimization}
\label{optimization}

Here we describe the estimation algorithm for fixed penalty parameters $\lambda_1$ and $\lambda_2$;  selection of these parameters is discussed in Section~\ref{tuning}. To obtain estimates of the coefficient array $\mathbb{B}$ and intercept $\b_0$, we iteratively optimize the objective function $h(\mathbf{y}, \mathbb{X}; \mathbb{B}, b_0)$ \eqref{eq_genmultdwd} for each dimension to obtain the estimated weights for that dimension with other dimension's weights fixed. This general iterative estimation approach is described in Algorithm~\ref{alg1}.

\begin{algorithm}[H]
\SetAlgoLined
\begin{enumerate}
\item \emph{Initialization.} Generate $K$ random matrices $\tilde{\mathbf{U}}_k: P_k \times R$ for $k=1, ..., K$. In our implementation,  entries are generated independently from a Uniform$[0,1]$ distribution. Compute the coefficient array $\tilde{\mathbb{B}}=\llbracket \mathbf{U}_1,\dots,\mathbf{U}_K \rrbracket$,  and initialize $\tilde{b}_0=0$.  
\item \emph{Iteration.} Update $\tilde{\U}_1$ and $\tilde{b}_0$ by optimizing the conditional objective,
\begin{align}
\{\tilde{b}_0, \tilde{\U}_1 \} = \underset{\{b_0, \U_1 \}}{\arg \min} \; \; h(\mathbf{y}, \mathbb{X}; \llbracket \mathbf{U}_1, \tilde{\mathbf{U}}_2,\dots,\tilde{\mathbf{U}}_K \rrbracket, b_0).
\label{sub-step}
\end{align}
Similarly update $\tilde{\U}_2,\hdots,\tilde{\U}_K$, with the value for $b_0$ updated at each step.  The details of the optimization sub-step~\eqref{sub-step} are described below.
\item \emph{Convergence}. Set $\mathbb{B}^{\text{New}}=\llbracket \tilde{\mathbf{U}}_1,\dots,\tilde{\mathbf{U}}_K \rrbracket$.  If $||\mathbb{B}^{\text{New}}-\tilde{\mathbb{B}}||_2^2>\epsilon$ for some pre-specified threshold $\epsilon$, update $\tilde{\mathbb{B}}=\mathbb{B}^{\text{New}}$ and repeat Step 2. Otherwise, our final estimates are $\hat{\BB}=\tilde{\BB}$ and $\hat{b}_0=\tilde{b}_0$.
\end{enumerate}
 \caption{General estimation steps.}
 \label{alg1}
\end{algorithm}

The procedure for solving the optimization problem in equation \eqref{sub-step} depends on whether the model is rank-$1$ $(R=1)$ or higher rank ($R \geq 2$).  If $R=1$, then $\BB=\u_1 \circ \cdots \circ \u_2$ and the objective in~\eqref{sub-step} can be expressed as 
$$\frac{1}{N} \sum_{i=1}^{N} V \left(y_{i} \left(b_0+\tilde{\x}_i^{(1) T} \u_1  \right) \right) + \left(\lambda_1 \prod_{k=2}^K \|\tilde{\mathbf{u}}_k\|_1\right)\|\mathbf{u}_1\|_1+ \left(\frac{\lambda_2}{2} \prod_{k=2}^K \|\tilde{\mathbf{u}}_k\|^2_2\right)\|\mathbf{u}_1\|_2^2,$$
where 
\begin{align}
\tilde{\x}_i^{(1)}[p_1]=\XX_i[p_1,\dots]\cdot (\tilde{\mathbf{u}}_2 \circ \cdots \circ \tilde{\mathbf{u}}_K) \; \text{ for }  p_1=1,\hdots,P_1.
\label{xi1}
\end{align}
This is equivalent to the vector sparse DWD objective \eqref{eq_dwd3} as a function of $b_0$ and $\u_1$, and thus can be solved using existing software such as the {\tt sdwd} function in the R package {\tt SDWD} \citep{sdwd}.     
For rank $R\geq2$, we solve~\eqref{sub-step} using a coordinate descent algorithm based on the majorization-minimization (MM) principle \citep{hunter2004tutorial}, thus extending the algorithm described in \cite{wang2016sparse}.  
The steps for this procedure are given in Algorithm~\ref{alg2}, and analogous updates are used for $\U_2,\hdots,\U_K$.  

To derive Step b of Algorithm~\ref{alg2}, consider replacing $\tilde{\U}_1[j,r]$ with $\U_1[j,r]$ in $\tilde{\BB}$ to obtain $\BB^*$.  Then, 
\begin{align*} h(\mathbf{y}, \mathbb{X}; \mathbb{B}^*, b_0)=  \frac{1}{N} \sum_{i=1}^{N} V \left(\tilde{\mu}_i+y_i \tilde{\X}_i^{(1)}[j,r] (\U_1[j,r]-\tilde{\U}_1[j,r])\right)+P_{\lambda_{1}, \lambda_{2}}(\mathbb{B}^*), \end{align*}
and we approximate $h(\mathbf{y}, \mathbb{X}; \mathbb{B}^*, b_0)$ with a quadratic form analogous to that in \citet{wang2016sparse}, 
\begin{align} \frac{1}{N} \sum_{i=1}^N V(\mu_i) + \frac{1}{N} \sum_{i=1}^N V'\left(\tilde{\mu}_i\right) (\U_1[j,r]-\tilde{\U}_1[j,r])+2(\U_1[j,r]-\tilde{\U}_1[j,r])^2+P_{\lambda_{1}, \lambda_{2}}(\mathbb{B}^*).
\label{surrogate} 
\end{align}
Note that 
$$P_{\lambda_{1}, \lambda_{2}}(\mathbb{B}^*) = \lambda_1 q_r |\U_1[j,r]|+ \frac{\lambda_2}{2} \left( \W[r,r] \U_1[j,r]^2+ \sum_{r'\neq r}^R \W[r,r']\tilde{\U}_1[j,r'] \right)+C,$$
for a value $C$ that is constant with respect to $\U_1[j,r]$. Thus, the minimizer of \eqref{surrogate} over $\U_1[j,r]$ is given by $\tilde{\U}_1[j,r]^{\text{new}}$ in Step b.  A similar justification is used for the update of the intercept in Step c.  

\begin{algorithm}[H]
\SetAlgoLined
{\bf Step a:} Compute $\tilde{\X}_i^{(1)}: P_1 \times R$ with columns defined as in~\eqref{xi1} for each rank component $r=1,\hdots,R$: 
\begin{equation*}
\tilde{\X}_i^{(1)}[p_1,r]=\XX_i[p_1,\dots]\cdot (\tilde{\mathbf{u}}_{2r} \circ \cdots \circ \tilde{\mathbf{u}}_{Kr})    
\end{equation*}
and let $\tilde{\mu}_i = y_i(\tilde{b}_0+\tilde{\U}_1 \cdot \tilde{\X}_i^{(1)})$ for $i=1,\hdots,N$ (note: $\tilde{\U}_1 \cdot \tilde{\X}_i^{(1)}=\XX_i \cdot \tilde{\BB}$).  Define $\W: R \times R$ by $\W=(\U_2^T \U_2) \cdots (\U_K^T \U_K)$.

{\bf Step b:}  Update $\mathbf{U}_1$ via the MM principle and cyclic coordinate decent over $j=1,\dots,P_1$ and $r=1,\dots,R$. 

\begin{itemize}
    \item Compute $z= 4\tilde{\U}_1[j,r]-\frac{1}{N}\sum_{i=1}^N V'(\tilde{\mu}_i)\tilde{\X}_i^{(1)}[j,r] y_i - \lambda_2\sum_{r'\neq r}^R \W[r,r']\tilde{\U}_1[j,r']$, where $V'$ is the derivative of $V$ in \eqref{eq_dwd2}.  
    \item Compute $\tilde{\U}_1[j,r]^{\text{new}}=\frac{S(z, \lambda_1 q_r)}{4+\lambda_2 \W[r,r]}$, where $q_r=\prod_{k=2}^K \|\tilde{U}_{kr}\|_1$ is the weight for the $L_1$ penalty, and $S(z,g) = sign(z)(|z|-g)_{+}$ is the soft-thresholding operator in which $\omega_{+} = max(\omega,0)$. 
    \item Update $\tilde{\mu}_i = \tilde{\mu_i} +y_i(\tilde{\U}_1[j.r]^{\text{new}}-\tilde{\U}_1[j,r]) \tilde{\mathbf{X}}_{i}[j,r]$ for $i=1,\hdots,N$.
    \item Set $ \tilde{\U}_1[j,r]= \tilde{\U}_1[j,r]^{\text{new}}$. 
\end{itemize}
{\bf Step c:} Update the intercept. 
\begin{itemize}
    \item Compute $\tilde{b}_0^{\text{new}}=\tilde{b}_0 - \sum_{i=1}^N V'(\tilde{\mu}_i)y_i/(4N)$
    \item Update $\tilde{\mu}_i = \tilde{\mu}_i + y_i(\tilde{b}_0^{\text{new}}-\tilde{b}_0)$
    \item Set $\tilde{b}_0=\tilde{b}_0^{\text{new}}$.
\end{itemize}
{\bf Step d:} Repeat steps b-c until the difference between new estimates and previous estimates is smaller than a prespecified threshold, for example, $\sum_{j=1}^{P_1} \sum_{r=1}^R(\tilde{\U}_1[j,r]^{\text{new}}-\tilde{\U}_1[j,r])^2+(\tilde{b}_0^{\text{new}}-\tilde{b}_0)^2 < \epsilon$. At the end of iterations, we have the updated weight matrix $\mathbf{U}_1$.
 \caption{Update for $\U_1$ when $R \geq 2$.}
 \label{alg2}
\end{algorithm}

In general, Algorithm~\ref{alg1} does not guarantee convergence to a global optimum, and it can converge to local optima. We conducted simulation studies to assess the convergence of the proposed method and other alternative methods in Section~\ref{sim:convergence}, and in Appendix~\ref{app_convergence}. In practice, we find that starting with multiple initial values and pruning the paths with inferior objective values can improve the algorithm and obtain more robust results.

\subsection{Selection of tuning parameters} 
\label{tuning}

To determine the best pair of $\lambda_1$ and $\lambda_2$, K-fold cross-validation can be used across a grid of $\lambda_1$ and $\lambda_2$ values.  For example, by default we set the following candidates for the tuning parameters: $\lambda_1 = (10^{-4},0.001,0.005,0.01,0.025,0.05,0.1, 0.25, 0.5, 0.75, 1)$ and $\lambda_2=(0.25,0.50,0.75,1.00,3,5)$. The misclassification rate under cross-validation is one potential criterion; however, different parameter values may yield the same misclassification rate, especially if the classes are perfectly separated or nearly perfectly separated (see Section S7).  We thus recommend using t-test statistics for the test-data DWD scores between the two classes as a more general measure of separation for selecting tuning parameters. For each pair of $\lambda_1$ and $\lambda_2$ we compute the predicted DWD scores for all subjects via K-fold cross-validation, and then report t-test statistics for the difference of the predicted DWD scores between the two classes using the test-data. We selected as optimal the pair of $\lambda_1$ and $\lambda_2$ with the maximum t-test statistic.  For a fixed $\lambda_2$, the warm-start trick is used in the algorithm to select the optimal $\lambda_1$. That is, we use the solution at a smaller $\lambda_1$ as the initial value (the warm-start) to compute the solution at the next $\lambda_1$, improving computational efficiency. The performance of cross-validation for selecting parameters is evaluated through simulation studies presented in Appendix~\ref{app_cv}.

\section{Simulations}
\label{sec:simulations}

In our simulation studies, we first compare the proposed rank-1 multiway sparse model with the existing methods in Section~\ref{sim:rank1}. In Section~\ref{sim:convergence} we explore convergence and its impact on performance. In Section~\ref{sim:rankR}, we show the rank-R sparse model performs well when the true data generating model has a higher rank.  Additional simulation studies are presented in the appendices, including simulations to assess cross-validation for parameter selection, rank mispecification, correlated predictors, and different penalization approaches.  

\subsection{Rank-1 model simulation design}
\label{sim:rank1}

To evaluate the performance of the rank-1 multiway sparse DWD model (M-SDWD), we compared it with a rank-1 multiway sparse DWD model with $\lambda_1=0$ (M-SDWD $\lambda_1=0$), non-sparse multiway DWD 
(M-DWD), and the full model sparse DWD (Full SDWD). For M-DWD we use an extended version of multiway DWD \citep{lyu2017discriminating} that allows for data of any dimension $K>2$. The full model sparse DWD is a naive way to analyze multiway data by applying the standard sparse DWD \citep{wang2016sparse} to the vectorized multiway data, without considering multiway structure (i.e., no rank constraint). In this simulation study, data were generated under several conditions, including different multiway array dimensions, sample sizes, and sparsity levels. For all scenarios, training datasets with two classes of equal size $(N_0 = N_1 = N/2, N= 40 \; \text{or} \; 100)$ were generated. The predictors have the form of a three-way array of dimensions $\XX_i: P_1\times P_2\times P_3$. We consider two settings of different dimensionality: higher dimensional ($30 \times 15\times 15$) and lower dimensional ($15\times 4\times 5$). In each training dataset, for the $N_0$ samples corresponding to class -$1$, the entries of $\XX_i$ were generated independently from a $N(0,1)$ distribution. 
For the other $N_1$ samples corresponding to class $1$, the entries of $\XX_i$ were generated independently from a normal distribution with variance $1$ and the mean for each entry given by the array $\sqrt{\alpha}\mu_1$ where $\mu_1= \u_1 \circ \u_2 \circ \u_3$; the non-zero values of $\u_k, k = 1, 2, 3,$ were generated independently from a $N(0, 1)$ distribution. Here, $\alpha$ gives the signal to noise ratio, i.e., the variance of the mean difference between the two classes over the residual variance.  We set $\alpha = 0.2$ for this simulation study. The appendices present results with different signal to noise ratios (Appendix~\ref{app_s2n}) and with correlated predictors (Appendix~\ref{app_corr}).

We consider three scenarios with varying degrees of sparsity: more sparsity, less sparsity and no sparsity. Under the more sparsity scenario, about $1/3$ of the mean weights in each dimension ($\{\u_1, \u_2, \u_3\}$)  are set to zero. Specifically, under the high dimensional case ($30 \times 15\times 15$), 25 entries of $\u_1$, and 10 entries each of $\u_2$ and $\u_3$ are set to zero, that is, $5 \times 5\times 5$ of the variables have signal discriminating the classes; under the low dimensional case ($15 \times 4 \times 5$), only 5 entries of $\u_1$, and 2 entries each of $\u_2$ and $\u_3$ are non-zero. Under the less sparsity scenario, sparsity is considered for only one dimension. In particular, 10 entries of $\u_1$ under both the high and low dimensional cases are set to zero, and the rest of the $\u_k$ are non-zero. Under the no sparsity case, all variables for each $\u_k$ are nonzero. For the high dimensional case we also considered an additional case with even more sparsity in the model: only 3 variables of each dimension have signals and the rest of variables for each dimension are zero. Each scenario was replicated 200 times.  

Under all these scenarios, we computed the correlation of the estimated hyperplane and the true hyperplane. Here the ``true" hyperplane is the mean difference between the classes $\mu_1$, which corresponds to the Bayes linear classifier. 
We assess predictive performance by considering misclassification rates for test data that were generated from the same distributions as the training data with the same sample sizes ($N = 40$ or $100$). To assess recovery of the sparsity structure, we computed the true positive rates and true negative rates for the proportions of non-zero/zero weights that were correctly estimated. All these statistics were computed based on the vectorized weights for multi-way based methods. The margin of error across the replicates for each statistic was also computed.

Table \ref{tab_sim_low} and Table \ref{tab_sim_high} summarize the simulation results for lower and higher dimensional data, respectively. Both tables show the proposed rank-1 multiway sparse DWD model has the best performance, with higher correlation and lower misclassification rates than other methods, when the true model is more sparse. When the true model is less sparse or not sparse, the proposed method has comparable performances to multiway DWD (M-DWD). All multiway based methods have higher correlations and lower misclassification rates than the full sparse DWD model under different scenarios. Overall the proposed rank-1 multiway sparse DWD model performs well, but some of the advantages are not obvious. In particular, the mean correlations are relatively low for the very sparse models, since the correlations with the truth are very small for some simulation replications. This is because the algorithm may converge to local minima, which is explored further in Section~\ref{sim:convergence}. 

\begin{table}
\centering
\caption{Simulation results under the low dimensional scenario ($15 \times 4 \times 5$). In the Sparsity column, the numbers in  parentheses indicate the number of non-zero variables in each dimension. ``Cor'' is the correlation between the estimated linear hyperplane and the true hyperplane. ``Mis'' is the average misclassification rate. ``TP'' is the true positive rate, i.e., the proportion of non-zero coefficients that are correctly estimated to be non-zero.``TN'' is the true negative rate, i.e., the proportion of zero coefficients that are correctly estimated to be zero. The margins of error (2* standard deviations across 200 replicates) for each statistic are also listed following the $\pm$ symbol.}
\scalebox{0.8}{
\begin{tabular}{lllllll}
  \hline
N & Sparsity & Methods & Cor & Mis & TP & TN \\ 
  \hline
40 & More ($5 \times 2 \times 2$) & M-SDWD & \bf{0.359$\pm$0.054} & 0.390$\pm$0.025 & 0.492$\pm$0.057 & 0.644$\pm$0.056 \\ 
    &   & M-SDWD ($\lambda_1=0$) & 0.318$\pm$0.049 & \bf{0.386$\pm$0.025} & 1.000$\pm$0.000 & 0.000$\pm$0.000 \\ 
    &   & M-DWD & 0.313$\pm$0.049 & 0.388$\pm$0.026 & 1.000$\pm$0.000 & 0.000$\pm$0.000 \\ 
    &   & Full SDWD & 0.284$\pm$0.039 & 0.402$\pm$0.023 & 0.337$\pm$0.041 & 0.748$\pm$0.043 \\ 
    & Less ($5 \times 4 \times 5$) & M-SDWD & \bf{0.773$\pm$0.034} & 0.156$\pm$0.024 & 0.646$\pm$0.045 & 0.527$\pm$0.053 \\ 
    &   & M-SDWD ($\lambda_1=0$) & 0.768$\pm$0.035 & \bf{0.146$\pm$0.022} & 1.000$\pm$0.000 & 0.000$\pm$0.000 \\ 
    &   & M-DWD & 0.757$\pm$0.042 & 0.154$\pm$0.024 & 1.000$\pm$0.000 & 0.000$\pm$0.000 \\ 
    &   & Full SDWD & 0.532$\pm$0.031 & 0.214$\pm$0.024 & 0.320$\pm$0.038 & 0.776$\pm$0.039 \\ 
    & No ($15 \times 4 \times 5$)& M-SDWD & 0.894$\pm$0.025 & 0.052$\pm$0.015 & 0.792$\pm$0.038 &  - \\ 
    &   & M-SDWD ($\lambda_1=0$) & \bf{0.901$\pm$0.026} & \bf{0.048$\pm$0.014} & 1.000$\pm$0.000 &  - \\ 
    &   & M-DWD & 0.896$\pm$0.030 & 0.054$\pm$0.017 & 1.000$\pm$0.000 &  - \\ 
    &   & Full SDWD & 0.639$\pm$0.026 & 0.096$\pm$0.020 & 0.444$\pm$0.045 &  - \\ 
  100 & More ($5 \times 2 \times 2$)& M-SDWD & \bf{0.574$\pm$0.056} & 0.332$\pm$0.025 & 0.536$\pm$0.055 & 0.713$\pm$0.051 \\ 
    &   & M-SDWD ($\lambda_1=0$) & 0.489$\pm$0.055 & \bf{0.329$\pm$0.025} & 1.000$\pm$0.000 & 0.000$\pm$0.000 \\ 
    &   & M-DWD & 0.489$\pm$0.056 & 0.331$\pm$0.025 & 1.000$\pm$0.000 & 0.000$\pm$0.000 \\ 
    &   & Full SDWD & 0.448$\pm$0.049 & 0.337$\pm$0.024 & 0.318$\pm$0.037 & 0.841$\pm$0.034 \\ 
    & Less ($5 \times 4 \times 5$) & M-SDWD & \bf{0.902$\pm$0.022} & 0.112$\pm$0.018 & 0.720$\pm$0.038 & 0.551$\pm$0.050 \\ 
    &   & M-SDWD ($\lambda_1=0$) & 0.890$\pm$0.025 & 0.113$\pm$0.018 & 1.000$\pm$0.000 & 0.000$\pm$0.000 \\ 
    &   & M-DWD & 0.898$\pm$0.027 & \bf{0.111$\pm$0.018} & 1.000$\pm$0.000 & 0.000$\pm$0.000 \\ 
    &   & Full SDWD & 0.695$\pm$0.028 & 0.156$\pm$0.021 & 0.357$\pm$0.034 & 0.794$\pm$0.033 \\ 
    & No ($15 \times 4 \times 5$)& M-SDWD & 0.954$\pm$0.014 & \bf{0.035$\pm$0.012} & 0.870$\pm$0.031 &  - \\ 
    &   & M-SDWD ($\lambda_1=0$) & 0.963$\pm$0.012 & \bf{0.035$\pm$0.011} & 1.000$\pm$0.000 &  - \\ 
    &   & M-DWD & \bf{0.965$\pm$0.014} & \bf{0.035$\pm$0.012} & 1.000$\pm$0.000 &  - \\ 
    &   & Full SDWD & 0.782$\pm$0.021 & 0.061$\pm$0.015 & 0.433$\pm$0.036 &  - \\ 
   \hline
\end{tabular}}
\label{tab_sim_low}
\end{table}

\begin{table}
\centering
\caption{Simulation results under the high dimensional scenario ($30 \times 15 \times 15$). In the Sparsity column, the numbers in  parentheses indicate the number of non-zero variables in each dimension. ``Cor'' is the correlation between the estimated linear hyperplane and the true hyperplane. ``Mis'' is the average misclassification rate. ``TP'' is the true positive rate, i.e., the proportion of non-zero coefficients that are correctly estimated to be non-zero.``TN'' is the true negative rate, i.e., the proportion of zero coefficients that are correctly estimated to be zero. The margins of error (2* standard deviations across 200 replicates) for each statistic are also listed following the $\pm$ symbol.}
\scalebox{0.8}{
\begin{tabular}{lllllll}
  \hline
N & Sparsity & Methods & Cor & Mis & TP & TN \\ 
  \hline
40 & Even more ($3 \times 3 \times 3$) & M-SDWD & 0.214$\pm$0.053 & 0.426$\pm$0.025 & 0.077$\pm$0.012 & 0.738$\pm$0.054 \\ 
    &   & M-SDWD ($\lambda_1=0$) & 0.154$\pm$0.042 & 0.415$\pm$0.025 & 0.216$\pm$0.000 & 0.000$\pm$0.000 \\ 
    &   & M-DWD & 0.171$\pm$0.043 & 0.404$\pm$0.025 & 0.216$\pm$0.000 & 0.000$\pm$0.000 \\ 
    &   & Full SDWD & \bf{0.233$\pm$0.041} & \bf{0.388$\pm$0.024} & 0.039$\pm$0.008 & 0.900$\pm$0.035 \\ 
    & More ($5 \times 5 \times 5$) & M-SDWD & \bf{0.644$\pm$0.056} & 0.218$\pm$0.033 & 0.489$\pm$0.052 & 0.778$\pm$0.047 \\ 
    &   & M-SDWD ($\lambda_1=0$) & 0.516$\pm$0.056 & \bf{0.216$\pm$0.031} & 1.000$\pm$0.000 & 0.000$\pm$0.000 \\ 
    &   & M-DWD & 0.505$\pm$0.058 & 0.236$\pm$0.033 & 1.000$\pm$0.000 & 0.000$\pm$0.000 \\ 
    &   & Full SDWD & 0.389$\pm$0.039 & 0.238$\pm$0.027 & 0.129$\pm$0.026 & 0.932$\pm$0.026 \\ 
    & Less ($10 \times 15 \times 15$) & M-SDWD & 0.983$\pm$0.002 & 0.000$\pm$0.000 & 0.872$\pm$0.026 & 0.394$\pm$0.060 \\ 
    &   & M-SDWD ($\lambda_1=0$) & 0.988$\pm$0.001 & 0.000$\pm$0.000 & 1.000$\pm$0.000 & 0.000$\pm$0.000 \\ 
    &   & M-DWD & \bf{0.990$\pm$0.001} & 0.000$\pm$0.000 & 1.000$\pm$0.000 & 0.000$\pm$0.000 \\ 
    &   & Full SDWD & 0.654$\pm$0.016 & 0.002$\pm$0.002 & 0.143$\pm$0.019 & 0.943$\pm$0.017 \\ 
    & No ($30 \times 15 \times 15$)& M-SDWD & 0.987$\pm$0.003 & 0.000$\pm$0.000 & 0.869$\pm$0.028 &  - \\ 
    &   & M-SDWD ($\lambda_1=0$) & 0.994$\pm$0.001 & 0.000$\pm$0.000 & 1.000$\pm$0.000 &  - \\ 
    &   & M-DWD & \bf{0.997$\pm$0.000} & 0.000$\pm$0.000 & 1.000$\pm$0.000 &  - \\ 
    &   & Full SDWD & 0.724$\pm$0.010 & 0.000$\pm$0.000 & 0.262$\pm$0.032 &  - \\ 
  100 & Even more ($3 \times 3 \times 3$)& M-SDWD & 0.140$\pm$0.045 & 0.467$\pm$0.016 & 0.018$\pm$0.004 & 0.784$\pm$0.049 \\ 
    &   & M-SDWD ($\lambda_1=0$) & 0.084$\pm$0.032 & 0.464$\pm$0.016 & 0.064$\pm$0.000 & 0.000$\pm$0.000 \\ 
    &   & M-DWD & 0.101$\pm$0.035 & 0.455$\pm$0.017 & 0.064$\pm$0.000 & 0.000$\pm$0.000 \\ 
    &   & Full SDWD & \bf{0.218$\pm$0.047} & \bf{0.431$\pm$0.020} & 0.018$\pm$0.003 & 0.868$\pm$0.038 \\ 
    & More ($5 \times 5 \times 5$) & M-SDWD & \bf{0.849$\pm$0.038} & \bf{0.089$\pm$0.020} & 0.668$\pm$0.040 & 0.806$\pm$0.041 \\ 
    &   & M-SDWD ($\lambda_1=0$) & 0.796$\pm$0.040 & 0.101$\pm$0.021 & 1.000$\pm$0.000 & 0.000$\pm$0.000 \\ 
    &   & M-DWD & 0.766$\pm$0.049 & 0.121$\pm$0.025 & 1.000$\pm$0.000 & 0.000$\pm$0.000 \\ 
    &   & Full SDWD & 0.636$\pm$0.035 & 0.136$\pm$0.022 & 0.172$\pm$0.023 & 0.957$\pm$0.022 \\ 
    & Less ($10 \times 15 \times 15$) & M-SDWD & 0.992$\pm$0.001 & 0.000$\pm$0.000 & 0.890$\pm$0.021 & 0.461$\pm$0.060 \\ 
    &   & M-SDWD ($\lambda_1=0$) & 0.995$\pm$0.000 & 0.000$\pm$0.000 & 1.000$\pm$0.000 & 0.000$\pm$0.000 \\ 
    &   & M-DWD & \bf{0.996$\pm$0.000} & 0.000$\pm$0.000 & 1.000$\pm$0.000 & 0.000$\pm$0.000 \\ 
    &   & Full SDWD & 0.816$\pm$0.009 & 0.000$\pm$0.001 & 0.199$\pm$0.014 & 0.954$\pm$0.008 \\ 
    & No ($30 \times 15 \times 15$)& M-SDWD & 0.995$\pm$0.001 & 0.000$\pm$0.000 & 0.934$\pm$0.016 &  - \\ 
    &   & M-SDWD ($\lambda_1=0$) & 0.997$\pm$0.001 & 0.000$\pm$0.000 & 1.000$\pm$0.000 &  - \\ 
    &   & M-DWD & \bf{0.999$\pm$0.000} & 0.000$\pm$0.000 & 1.000$\pm$0.000 &  - \\ 
    &   & Full SDWD & 0.848$\pm$0.005 & 0.000$\pm$0.000 & 0.342$\pm$0.027 &  - \\ 
   \hline
\end{tabular}}
\label{tab_sim_high}
\end{table}

\subsection{Assessment of convergence}
\label{sim:convergence}

\begin{figure}
    \centering
    \includegraphics[scale=0.9]{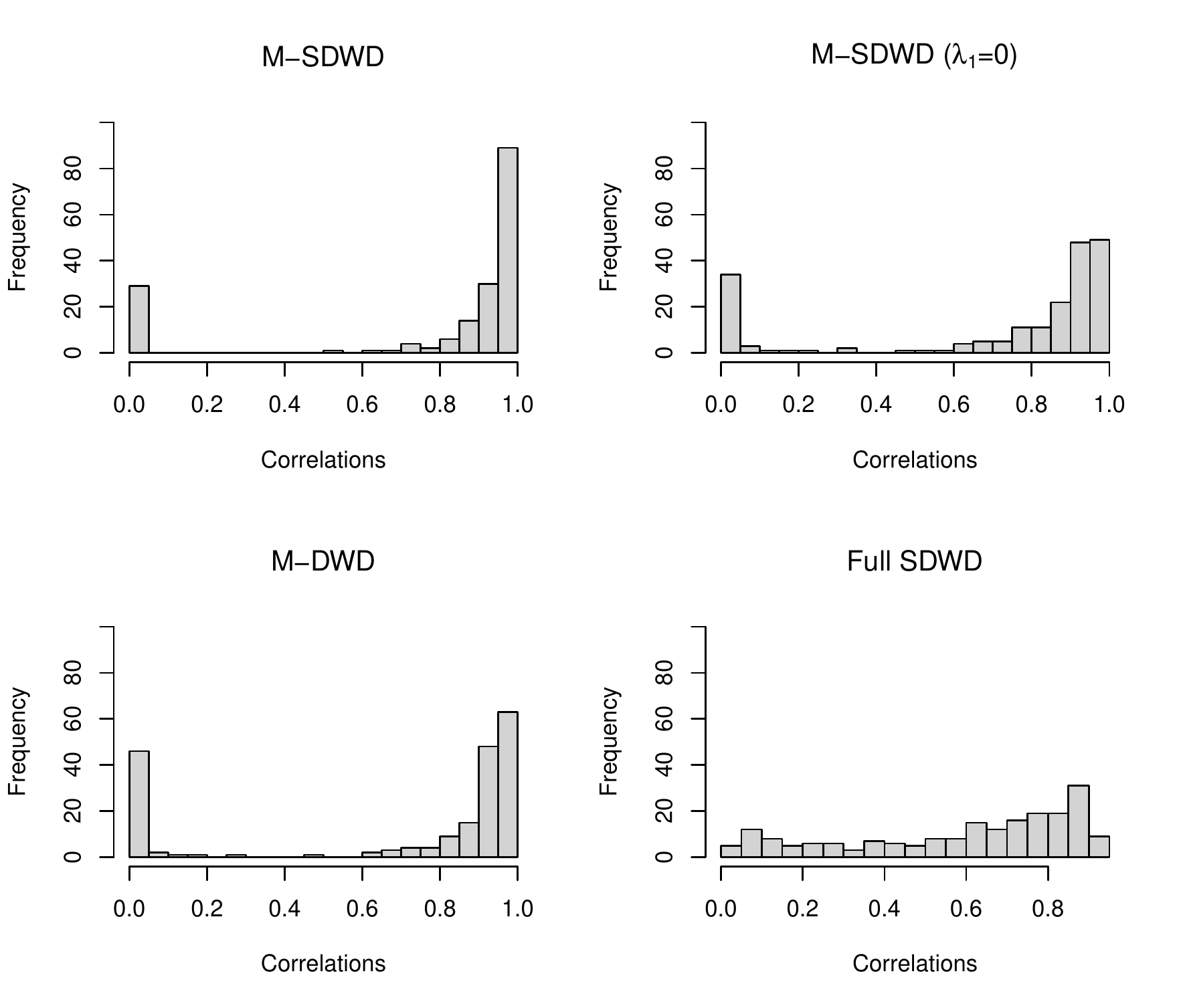}
    \caption{Histogram of correlations between true hyperplane and estimates with four classification methods under the more sparsity case based on high dimensional data $(30 \times 15 \times 15)$ and sample size $N =100$.}
\label{fig_hist_2}
\end{figure}

To assess the convergence of the four methods, we show the distributions of correlations between the true and estimated hyperplane for the ``more sparsity" scenario for high dimensional data in Figure \ref{fig_hist_2}. The distribution is bimodal for the three multiway methods, where for some replications the signal is estimated very well (correlation$\approx$1) and others do not recover the true signal at all (correlation$\approx$0).  We find that the poor performing replications are due to the algorithm converging to a local optimum, which does not occur for the full (vectoriozed) SDWD method. The proposed multiway sparse DWD (M-SDWD) performs better than other methods with more reasonable correlations. In addition, the bimodality becomes less severe when the signal to noise ratio increases (see Appendix~\ref{app_convergence}). Considering prediction performance without separating out the replications that do not converge appropriately may complicate interpretation of the results, hense we present another two tables that show those simulations with correlation greater than 0.5 for each method, and the statistics among those simulations with correlation greater than $0.5$ for that method. From Table \ref{tab_sim_high_5} and Table \ref{tab_sim_low_5} we can see the proposed method performs the best under the ``even more" and ``more" sparsity cases. To reduce convergence to local optima, we recommend running the algorithm with multiple initial values for the first several iterations and selecting the optimal path to proceed.

\begin{table}
\centering
\caption{Simulation results among simulations with correlation greater than 0.5 under the high dimensional scenario ($30 \times 15 \times 15$). In the Sparsity column, the numbers in  parentheses indicate the number of non-zero variables in each dimension. ``Cor'' is the correlation between the estimated linear hyperplane and the true hyperplane. ``Mis'' is the average misclassification rate. ``TP'' is the true positive rate, i.e., the proportion of non-zero coefficients that are correctly estimated to be non-zero.``TN'' is the true negative rate, i.e., the proportion of zero variables that are correctly estimated to be zero. The margins of error (2* standard deviations across 200 replicates) for each statistic are also listed following the $\pm$ symbol.}
\scalebox{0.7}{
\begin{tabular}{llllllll}
\hline
N & Sparsity & Methods & Cor & Mis & TP & TN & Prop. cor$>$0.5 \\ 
  \hline
40 & Even more ($3 \times 3 \times 3$) & M-SDWD & \bf{0.905$\pm$0.031} & \bf{0.045$\pm$0.018} & 0.145$\pm$0.020 & 0.853$\pm$0.109 & 0.16 \\ 
    &   & M-SDWD ($\lambda_1=0$) & 0.783$\pm$0.039 & 0.074$\pm$0.029 & 0.216$\pm$0.000 & 0.000$\pm$0.000 & 0.175 \\ 
    &   & M-DWD & 0.816$\pm$0.035 & 0.068$\pm$0.023 & 0.216$\pm$0.000 & 0.000$\pm$0.000 & 0.175 \\ 
    &   & Full SDWD & 0.739$\pm$0.031 & 0.112$\pm$0.031 & 0.033$\pm$0.006 & 1.014$\pm$0.001 & 0.22 \\ 
    & More ($5 \times 5 \times 5$) & M-SDWD & \bf{0.888$\pm$0.017} & \bf{0.034$\pm$0.011} & 0.638$\pm$0.052 & 0.802$\pm$0.055 & 0.61 \\ 
    &   & M-SDWD ($\lambda_1=0$) & 0.840$\pm$0.022 & 0.039$\pm$0.012 & 1.000$\pm$0.000 & 0.000$\pm$0.000 & 0.585 \\ 
    &   & M-DWD & 0.863$\pm$0.021 & 0.039$\pm$0.013 & 1.000$\pm$0.000 & 0.000$\pm$0.000 & 0.56 \\ 
    &   & Full SDWD & 0.682$\pm$0.021 & 0.044$\pm$0.014 & 0.091$\pm$0.015 & 0.997$\pm$0.001 & 0.41 \\ 
    & Less ($10 \times 15 \times 15$) & M-SDWD & 0.983$\pm$0.002 & 0.000$\pm$0.000 & 0.872$\pm$0.026 & 0.394$\pm$0.060 & 1 \\ 
    &   & M-SDWD ($\lambda_1=0$) & 0.988$\pm$0.001 & 0.000$\pm$0.000 & 1.000$\pm$0.000 & 0.000$\pm$0.000 & 1 \\ 
    &   & M-DWD & \bf{0.990$\pm$0.001} & 0.000$\pm$0.000 & 1.000$\pm$0.000 & 0.000$\pm$0.000 & 1 \\ 
    &   & Full SDWD & 0.682$\pm$0.013 & 0.000$\pm$0.000 & 0.135$\pm$0.015 & 0.956$\pm$0.011 & 0.885 \\ 
    & No ($30 \times 15 \times 15$)& M-SDWD & 0.987$\pm$0.003 & 0.000$\pm$0.000 & 0.869$\pm$0.028 &  - & 1 \\ 
    &   & M-SDWD ($\lambda_1=0$) & 0.994$\pm$0.001 & 0.000$\pm$0.000 & 1.000$\pm$0.000 &  - & 1 \\ 
    &   & M-DWD & \bf{0.997$\pm$0.000} & 0.000$\pm$0.000 & 1.000$\pm$0.000 &  - & 1 \\ 
    &   & Full SDWD & 0.727$\pm$0.009 & 0.000$\pm$0.000 & 0.262$\pm$0.032 &  - & 0.99 \\ 
  100 & Even more ($3 \times 3 \times 3$) & M-SDWD & \bf{0.922$\pm$0.051} & \bf{0.136$\pm$0.056} & 0.045$\pm$0.009 & 1.000$\pm$0.018 & 0.09 \\ 
    &   & M-SDWD ($\lambda_1=0$) & 0.794$\pm$0.064 & 0.146$\pm$0.056 & 0.064$\pm$0.000 & 0.000$\pm$0.000 & 0.085 \\ 
    &   & M-DWD & 0.784$\pm$0.060 & 0.153$\pm$0.042 & 0.064$\pm$0.000 & 0.000$\pm$0.000 & 0.115 \\ 
    &   & Full SDWD & 0.840$\pm$0.037 & 0.194$\pm$0.033 & 0.022$\pm$0.004 & 1.015$\pm$0.001 & 0.215 \\ 
    & More ($5 \times 5 \times 5$) & M-SDWD & \bf{0.935$\pm$0.010} & 0.048$\pm$0.010 & 0.697$\pm$0.038 & 0.831$\pm$0.042 & 0.905 \\ 
    &   & M-SDWD ($\lambda_1=0$) & 0.899$\pm$0.012 & 0.051$\pm$0.010 & 1.000$\pm$0.000 & 0.000$\pm$0.000 & 0.88 \\ 
    &   & M-DWD & 0.926$\pm$0.010 & \bf{0.042$\pm$0.009} & 1.000$\pm$0.000 & 0.000$\pm$0.000 & 0.825 \\ 
    &   & Full SDWD & 0.760$\pm$0.017 & 0.060$\pm$0.011 & 0.156$\pm$0.016 & 0.994$\pm$0.001 & 0.765 \\ 
    & Less ($10 \times 15 \times 15$) & M-SDWD & 0.992$\pm$0.001 & 0.000$\pm$0.000 & 0.890$\pm$0.021 & 0.461$\pm$0.060 & 1 \\ 
    &   & M-SDWD ($\lambda_1=0$) & 0.995$\pm$0.000 & 0.000$\pm$0.000 & 1.000$\pm$0.000 & 0.000$\pm$0.000 & 1 \\ 
    &   & M-DWD & \bf{0.996$\pm$0.000} & 0.000$\pm$0.000 & 1.000$\pm$0.000 & 0.000$\pm$0.000 & 1 \\ 
    &   & Full SDWD & 0.818$\pm$0.008 & 0.000$\pm$0.000 & 0.198$\pm$0.013 & 0.955$\pm$0.007 & 0.995 \\ 
    & No ($30 \times 15 \times 15$) & M-SDWD & 0.995$\pm$0.001 & 0.000$\pm$0.000 & 0.934$\pm$0.016 &  - & 1 \\ 
    &   & M-SDWD ($\lambda_1=0$) & 0.997$\pm$0.001 & 0.000$\pm$0.000 & 1.000$\pm$0.000 &  - & 1 \\ 
    &   & M-DWD & \bf{0.999$\pm$0.000} & 0.000$\pm$0.000 & 1.000$\pm$0.000 &  - & 1 \\ 
    &   & Full SDWD & 0.848$\pm$0.005 & 0.000$\pm$0.000 & 0.342$\pm$0.027 &  - & 1 \\ 
   \hline
\end{tabular}}
\label{tab_sim_high_5}
\end{table}

\begin{table}
\centering
\caption{Simulation results among simulations with correlation greater than 0.5 under the low dimensional scenario ($15 \times 4 \times 5$).  In the Sparsity column, the numbers in  parentheses indicate the number of non-zero variables in each dimension. ``Cor'' is the correlation between the estimated linear hyperplane and the true hyperplane. ``Mis'' is the average misclassification rate. ``TP'' is the true positive rate, i.e., the proportion of non-zero coefficients that are correctly estimated to be non-zero.``TN'' is the true negative rate, i.e., the proportion of zero coefficients that are correctly estimated to be zero. The margins of error (2* standard deviations across 200 replicates) for each statistic are also listed following the $\pm$ symbol.}
\scalebox{0.7}{
\begin{tabular}{llllllll}
  \hline
N & Sparsity & Methods & Cor & Mis & TP & TN & Prop. of Cor$>$0.5 \\ 
  \hline
40 & More (5 $\times$ 2 $\times$ 2) & M-SDWD & {\bf 0.829$\pm$0.028} & 0.159$\pm$0.030 & 0.616$\pm$0.077 & 0.768$\pm$0.079 & 0.31 \\ 
    &   & M-SDWD ($\lambda_1=0$) & 0.778$\pm$0.030 & 0.179$\pm$0.030 & 1.000$\pm$0.000 & 0.000$\pm$0.000 & 0.34 \\ 
    &   & M-DWD & 0.815$\pm$0.029 & {\bf 0.149$\pm$0.029} & 1.000$\pm$0.000 & 0.000$\pm$0.000 & 0.3 \\ 
    &   & Full SDWD & 0.721$\pm$0.038 & 0.159$\pm$0.033 & 0.277$\pm$0.052 & 0.955$\pm$0.019 & 0.22 \\ 
    & Less (5 $\times$ 4 $\times$ 5) & M-SDWD & 0.851$\pm$0.019 &  0.101$\pm$0.017 & 0.684$\pm$0.044 & 0.518$\pm$0.057 & 0.845 \\ 
    &   & M-SDWD ($\lambda_1=0$) & {\bf 0.853$\pm$0.018} & 0.102$\pm$0.017 & 1.000$\pm$0.000 & 0.000$\pm$0.000 & 0.865 \\ 
    &   & M-DWD & 0.877$\pm$0.017 & 0.096$\pm$0.016 & 1.000$\pm$0.000 & 0.000$\pm$0.000 & 0.84 \\ 
    &   & Full SDWD & 0.696$\pm$0.021 & {\bf 0.089$\pm$0.018} & 0.280$\pm$0.033 & 0.869$\pm$0.027 & 0.54 \\ 
    & No (15 $\times$ 4 $\times$ 5)& M-SDWD & 0.930$\pm$0.010 & 0.030$\pm$0.008 & 0.816$\pm$0.036 &  - & 0.95 \\ 
    &   & M-SDWD ($\lambda_1=0$) & 0.935$\pm$0.011 & 0.031$\pm$0.009 & 1.000$\pm$0.000 &  - & 0.96 \\ 
    &   & M-DWD & {\bf 0.947$\pm$0.010} & {\bf 0.027$\pm$0.009} & 1.000$\pm$0.000 &  - & 0.94 \\ 
    &   & Full SDWD & 0.724$\pm$0.016 & 0.029$\pm$0.008 & 0.434$\pm$0.049 &  - & 0.765 \\ 
  100 & More (5 $\times$ 2 $\times$ 2) & M-SDWD & {\bf 0.872$\pm$0.024} & 0.191$\pm$0.025 & 0.673$\pm$0.058 & 0.772$\pm$0.058 & 0.535 \\ 
    &   & M-SDWD ($\lambda_1=0$) & 0.850$\pm$0.023 &  0.186$\pm$0.024 & 1.000$\pm$0.000 & 0.000$\pm$0.000 & 0.515 \\ 
    &   & M-DWD & 0.853$\pm$0.026 & 0.190$\pm$0.025 & 1.000$\pm$0.000 & 0.000$\pm$0.000 & 0.525 \\ 
    &   & Full SDWD & 0.800$\pm$0.027 & {\bf 0.177$\pm$0.025} & 0.340$\pm$0.042 & 0.953$\pm$0.016 & 0.45 \\ 
    & Less (5 $\times$ 4 $\times$ 5) & M-SDWD & 0.929$\pm$0.011 & {\bf 0.090$\pm$0.014} & 0.749$\pm$0.033 & 0.536$\pm$0.051 & 0.945 \\ 
    &   & M-SDWD ($\lambda_1=0$) & 0.927$\pm$0.010 & 0.094$\pm$0.014 & 1.000$\pm$0.000 & 0.000$\pm$0.000 & 0.95 \\ 
    &   & M-DWD & {\bf 0.939$\pm$0.010} & 0.093$\pm$0.014 & 1.000$\pm$0.000 & 0.000$\pm$0.000 & 0.95 \\ 
    &   & Full SDWD & 0.768$\pm$0.018 & 0.106$\pm$0.016 & 0.336$\pm$0.032 & 0.838$\pm$0.028 & 0.83 \\ 
    & No (15 $\times$ 4 $\times$ 5)& M-SDWD & 0.962$\pm$0.009 & 0.031$\pm$0.010 & 0.879$\pm$0.030 &  - & 0.99 \\ 
    &   & M-SDWD ($\lambda_1=0$) & 0.968$\pm$0.008 & 0.033$\pm$0.010 & 1.000$\pm$0.000 &  - & 0.995 \\ 
    &   & M-DWD & {\bf 0.975$\pm$0.006} & {\bf 0.030$\pm$0.010} & 1.000$\pm$0.000 &  - & 0.99 \\ 
    &   & Full SDWD & 0.816$\pm$0.013 & 0.036$\pm$0.010 & 0.437$\pm$0.037 &  - & 0.925 \\ 
   \hline
\end{tabular}}
\label{tab_sim_low_5}
\end{table}

\subsection{Rank-R model simulation and results}
\label{sim:rankR}

In this simulation study, we aim to validate the higher rank multiway sparse model with a scenario in which $R=2$.  The procedure for data generation is similar to that of the rank-1 model. Instead of generating vectors $\u_k$ for each dimension, we generated three matrices $\U_k, k=1,2,3$ with dimensions $P_k \times 2$. The entries of $\U_k$ are generated independently from a normal distribution $N(0, 1)$, then we compute $\mu_1 = \llbracket \mathbf{U}_1,\dots,\mathbf{U}_K \rrbracket$. The $N_1$ samples corresponding to class $1$ were generated from a multivariate normal distribution $N(\mu_1, \mathbf{I}_{P_1P_2P_3 \times P_1P_2P_3})$. The $N_0$ samples in class -$1$ were generated independently from a $N(0,1)$ distribution. Different scenarios are considered in terms of dimensions, sample size and sparsity levels. The results in Table \ref{rank2tab} show that the rank-2 model has the best performance when the data were generated from a true rank-2 model under different scenarios. The performances of these methods are similar for a higher rank (R=5) and low dimensional case (see Appendix~\ref{app_highrank}).  

\begin{table}
\centering
\caption{Simulation results under the high dimensional scenario ($30 \times 15 \times 15$) when the true model is rank-2. In the Sparsity column, the numbers in  parentheses indicate the number of non-zero variables in each dimension. ``Cor'' is the correlation between the estimated linear hyperplane and the true hyperplane. ``Mis'' is the average misclassification rate. ``TP'' is the true positive rate, i.e., the proportion of non-zero coefficients that are correctly estimated to be non-zero.``TN'' is the true negative rate, i.e., the proportion of zero coefficients that are correctly estimated to be zero. The margins of error (2* standard deviations across 200 replicates) for each statistic are also listed following the $\pm$ symbol.}

\scalebox{0.7}{
\begin{tabular}{lllllll}
  \hline
N & Sparsity & Methods & Cor & Mis & TP & TN \\ 
  \hline
 40& More ($5 \times 5 \times 5$)   & M-SDWD (R=2) & {\bf 0.833$\pm$0.020} & {\bf 0.027$\pm$0.011} & 0.714$\pm$0.038 & 0.761$\pm$0.043 \\ 
   &   & M-SDWD ($\lambda_1=0$, R=2) & 0.747$\pm$0.025 & 0.040$\pm$0.013 & 1.000$\pm$0.000 & 0.000$\pm$0.000 \\ 
    &   & M-SDWD (R=1) & 0.779$\pm$0.025 & 0.035$\pm$0.014 & 0.678$\pm$0.039 & 0.716$\pm$0.050 \\ 
    &   & M-SDWD ($\lambda_1=0$, R=1) & 0.745$\pm$0.026 & 0.043$\pm$0.015 & 1.000$\pm$0.000 & 0.000$\pm$0.000 \\ 
    &   & M-DWD & 0.736$\pm$0.033 & 0.055$\pm$0.019 & 1.000$\pm$0.000 & 0.000$\pm$0.000 \\ 
    &   & Full SDWD & 0.591$\pm$0.024 & 0.071$\pm$0.015 & 0.150$\pm$0.013 & 0.984$\pm$0.007 \\ 
    & Less ($10 \times 15 \times 15$) & M-SDWD (R=2) & 0.984$\pm$0.003 & 0.000$\pm$0.000 & 0.942$\pm$0.016 & 0.402$\pm$0.060 \\ 
    &   & M-SDWD ($\lambda_1=0$, R=2) & {\bf 0.990$\pm$0.002} & 0.000$\pm$0.000 & 1.000$\pm$0.000 & 0.000$\pm$0.000 \\ 
    &   & M-SDWD (R=1) & 0.813$\pm$0.011 & 0.000$\pm$0.000 & 0.840$\pm$0.029 & 0.488$\pm$0.062 \\ 
    &   & M-SDWD ($\lambda_1=0$, R=1) & 0.821$\pm$0.010 & 0.000$\pm$0.000 & 1.000$\pm$0.000 & 0.000$\pm$0.000 \\ 
    &   & M-DWD & 0.786$\pm$0.016 & 0.000$\pm$0.000 & 1.000$\pm$0.000 & 0.000$\pm$0.000 \\ 
    &   & Full SDWD & 0.750$\pm$0.009 & 0.000$\pm$0.000 & 0.239$\pm$0.018 & 0.937$\pm$0.012 \\ 
    & No ($30 \times 15 \times 15$) & M-SDWD (R=2) & {\bf 0.996$\pm$0.000} & 0.000$\pm$0.000 & 1.000$\pm$0.000 &  - \\ 
    &   & M-SDWD ($\lambda_1=0$, R=2) & {\bf 0.996$\pm$0.000} & 0.000$\pm$0.000 & 1.000$\pm$0.000 &  - \\ 
    &   & M-SDWD (R=1) & 0.790$\pm$0.012 & 0.000$\pm$0.000 & 0.836$\pm$0.033 &  - \\ 
    &   & M-SDWD ($\lambda_1=0$, R=1) & 0.799$\pm$0.010 & 0.000$\pm$0.000 & 1.000$\pm$0.000 &  - \\ 
    &   & M-DWD & 0.763$\pm$0.016 & 0.000$\pm$0.000 & 1.000$\pm$0.000 &  - \\ 
    &   & Full SDWD & 0.784$\pm$0.006 & 0.000$\pm$0.000 & 0.399$\pm$0.036 &  - \\ 

100& More ($5 \times 5 \times 5$)  & M-SDWD (R=2) & {\bf 0.925$\pm$0.010} & {\bf 0.012$\pm$0.004} & 0.841$\pm$0.026 & 0.754$\pm$0.041 \\     &   & M-SDWD ($\lambda_1=0$, R=2) & 0.889$\pm$0.011 & 0.016$\pm$0.005 & 1.000$\pm$0.000 & 0.000$\pm$0.000 \\ 
    &   & M-SDWD (R=1) & 0.853$\pm$0.012 & 0.015$\pm$0.004 & 0.738$\pm$0.031 & 0.818$\pm$0.039 \\ 
    &   & M-SDWD ($\lambda_1=0$, R=1) & 0.829$\pm$0.014 & 0.019$\pm$0.005 & 1.000$\pm$0.000 & 0.000$\pm$0.000 \\ 
    &   & M-DWD & 0.825$\pm$0.020 & 0.027$\pm$0.010 & 1.000$\pm$0.000 & 0.000$\pm$0.000 \\ 
    &   & Full SDWD & 0.797$\pm$0.014 & 0.026$\pm$0.007 & 0.236$\pm$0.016 & 0.992$\pm$0.002 \\ 
    & Less ($10 \times 15 \times 15$)  & M-SDWD (R=2) & 0.992$\pm$0.001 & 0.000$\pm$0.000 & 0.958$\pm$0.012 & 0.456$\pm$0.061 \\ 
    &   & M-SDWD ($\lambda_1=0$, R=2) & {\bf 0.996$\pm$0.000} & 0.000$\pm$0.000 & 1.000$\pm$0.000 & 0.000$\pm$0.000 \\ 
    &   & M-SDWD (R=1) & 0.807$\pm$0.010 & 0.000$\pm$0.000 & 0.871$\pm$0.024 & 0.526$\pm$0.062 \\ 
    &   & M-SDWD ($\lambda_1=0$, R=1) & 0.807$\pm$0.010 & 0.000$\pm$0.000 & 1.000$\pm$0.000 & 0.000$\pm$0.000 \\ 
    &   & M-DWD & 0.797$\pm$0.012 & 0.000$\pm$0.000 & 1.000$\pm$0.000 & 0.000$\pm$0.000 \\ 
    &   & Full SDWD & 0.856$\pm$0.005 & 0.000$\pm$0.000 & 0.346$\pm$0.019 & 0.935$\pm$0.012 \\ 

    & No ($30 \times 15 \times 15$)  & M-SDWD (R=2) & {\bf 0.998$\pm$0.000} & 0.000$\pm$0.000 & 1.000$\pm$0.000 &  - \\ 
    &   & M-SDWD ($\lambda_1=0$, R=2) & {\bf 0.998$\pm$0.000} & 0.000$\pm$0.000 & 1.000$\pm$0.000 &  - \\ 
    &   & M-SDWD (R=1) & 0.786$\pm$0.009 & 0.000$\pm$0.000 & 0.908$\pm$0.023 &  - \\ 
    &   & M-SDWD ($\lambda_1=0$, R=1) & 0.790$\pm$0.009 & 0.000$\pm$0.000 & 1.000$\pm$0.000 &  - \\ 
    &   & M-DWD & 0.758$\pm$0.014 & 0.000$\pm$0.000 & 1.000$\pm$0.000 &  - \\ 
    &   & Full SDWD & 0.871$\pm$0.004 & 0.000$\pm$0.000 & 0.466$\pm$0.029 &  - \\ 
   \hline
\end{tabular}}
\label{rank2tab}
\end{table}

\section{Applications}
\label{sec:application}

\subsection{MRS data}
\label{mrs_app}

We apply the proposed method to our motivating application with MRS data to illustrate its utility. Friedreich's ataxia (FRDA) is an early-onset neurodegenerative disease caused by abnormalities in the frataxin gene \citep{10.1001/archneur.65.10.1296} resulting in knockdown of frataxin protein. The motivation of this study is to assess brain metabolic changes in a transgenic mouse model in which frataxin knockdown can be turned on by giving the antibiotic doxycycline.
Wild type (WT) mice and transgenic (TG) mice were randomly assigned to two different treatment groups: doxycycline treated group ($N_1=11$) and controls ($N_0=10$). Thus, the study included 4 groups of mice: WT treated with dox (WT-dox), WT without dox (WT-nodox), TG treated with dox (TG-dox), TG without dox (TG-nodox). Treatment was applied during weeks 1-12, then stopped during weeks 12-24 (recovery). Frataxin knockdown during weeks 1-12 was expected only in the TG-dox group, but not in WT animals nor in animals not receiving dox. The concentrations of 13 metabolites were measured in three different regions (cerebellum, cortex, and cervical spine). These animals were scanned at three time points: 0 weeks, 12 weeks and 24 weeks. The data have a multiway structure with four dimensions: mice $\times$ metabolites $\times$ regions $\times$ time, $\XX: 21 \times 13 \times 3 \times 3$. Our goal is to summarize the signal distinguishing the two treatment groups (dox vs. no dox) across the different dimensions, and assess how the effect of treatment depends on WT/TG status.  

\begin{figure}
    \centering
    \includegraphics[scale =0.7]{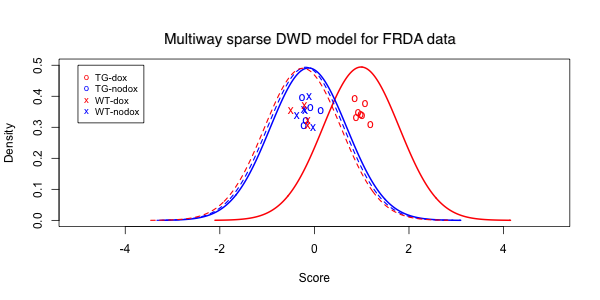}
    \caption{FRDA mouse study: Rank-1 multiway sparse DWD scores under leave-one-out cross-validation to classify mice that did or did not receive dox treatment.  The transgenic (TG) mice that received dox are clearly distinguished from the TG mice that did not and from the wild-type (WT) mice. A kernel density estimate is shown for the scores in each subgroup.}
\label{fig_sdwd_cv}
\end{figure}

The penalty parameters for multiway sparse DWD were selected through 10-fold cross validation (see Appendix~\ref{app_tuning}). Using the selected tuning parameters, Figure \ref{fig_sdwd_cv} shows the DWD scores computed under leave-one-out cross-validation based on the rank-1 multiway sparse DWD model.  These scores show robust separation of the no-dox and dox groups (which correspond to the two classes $y_i=-1$ and $y_i=1$) but only for the transgenic (TG) mice. The three scores for the TG-nodox, WT-nodox, and WT-dox groups are very similar. Because we train the classifier on both genetic cohorts but only observe a difference in the transgenic mice, this confirms the hypothesis that wild type mice are not affected by treatment; further, the results illustrate the robustness of the approach to within-class heterogeneity.

\begin{figure}
    \centering
    \includegraphics[scale =0.7]{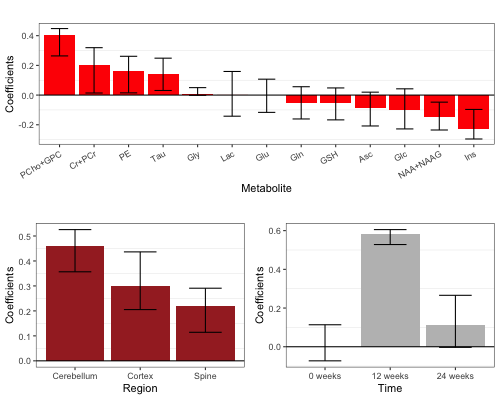}
    \caption{FRDA mouse study: Rank-1 multiway sparse DWD weights for metabolites, regions and time with $95\%$ bootstrap confidence intervals.}
    \label{fig_sdwd_load}
\end{figure}

In order to capture the uncertainty of the estimated weights and construct $95\%$ confidence intervals, 500 bootstrap samples were generated. For each bootstrap sample, 21 mice were resampled with replacement, and then the model was fit to the bootstrap sample to estimate weights for each dimension. The $2.5\%$ and $97.5\%$ quantiles 
for each estimated weight across the 500 bootstrap samples were computed to construct the $95\%$ confidence interval. Figure \ref{fig_sdwd_load} shows the estimated weights and their $95\%$ bootstrap confidence intervals. The metabolites with large absolute weights, such as PCho+GPC, Cr+PCr, and Ins, are considered important for Friedreich's ataxia research \citep{francca2009combined,iltis20101h,gramegna2017combined}, so were expected to have high weights for distinguishing the two groups. Two metabolites (Lac and Glu) and one time point (0 weeks) did not inform the classification as their estimated weights are exactly 0 in the full data fit.  Having no effect at baseline (0 weeks) makes sense as this is prior to the dox treatment. The observed changes at the later time points are similar to those observed in the R6/2 mouse, a severe mouse model of Huntington's disease \citep{zacharoff2012cortical}. This is consistent with the fact that both Friedreich's ataxia and Huntington's disease are characterized by impairment in energy metabolism.  

We compared misclassification rates for the M-SDWD, M-SDWD with $\lambda_1=0$, M-DWD, CATCH, Full SDWD and Random Forests by 10-fold cross validation. The 21 mice were randomly partitioned into 10 test subgroups of approximately equal size. 
This procedure was repeated $100$ times with different random partitions and the average test misclassification rates for dox vs. no dox were as follows: $22.1\%$ for M-SDWD, $20.9\%$ for M-SDWD with $\lambda_1=0$, $23.6\%$ for M-DWD,  $22.3\%$ for CATCH, $42.9\%$ for Full SDWD, and $22.3\%$ for Random Forest. To better reflect the reality of the treatment effect, we further considered the misclassification rate for TG-dox vs. the other three groups under this approach:
$3.1\%$ for M-SDWD, $1.9\%$ for M-SDWD with $\lambda_1=0$, $4.6\%$ for M-DWD,  $4.0\%$ for CATCH, $24.6\%$ for Full SDWD, and $10.5\%$ for Random Forest. Overall, the multiway methods perform better than the non-multiway methods (Full SDWD and Random Forest), illustrating the advantages of accounting for multiway structure. Moreover, the factorization of weights across dimensions in Figure~\ref{fig_sdwd_load} is useful for interpretation, and is not provided by the CATCH model. We conclude that the performance of M-SDWD model is competitive with alternatives. The rank-2 multiway sparse model was also applied, resulting in a  higher misclassification rate ($29.9\%$ for dox vs. no dox and $10.8\%$ for TG-dox vs. the other three groups), indicating that the rank-1 model is sufficient.    


\subsection{Gene course time data}

\begin{figure}
    \centering
    \includegraphics[scale =0.75]{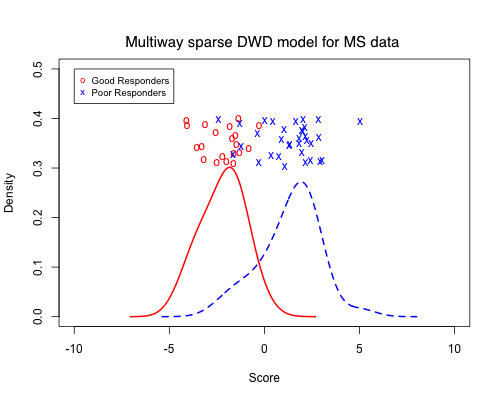}
    \caption{MS treatment study: Rank-1 multiway sparse DWD scores under leave-one-out cross-validation for good and poor treatment responders, with a kernel density estimate for each group. }
    \label{genefig1}
\end{figure}

\begin{figure}
    \centering
    \includegraphics[scale =0.75]{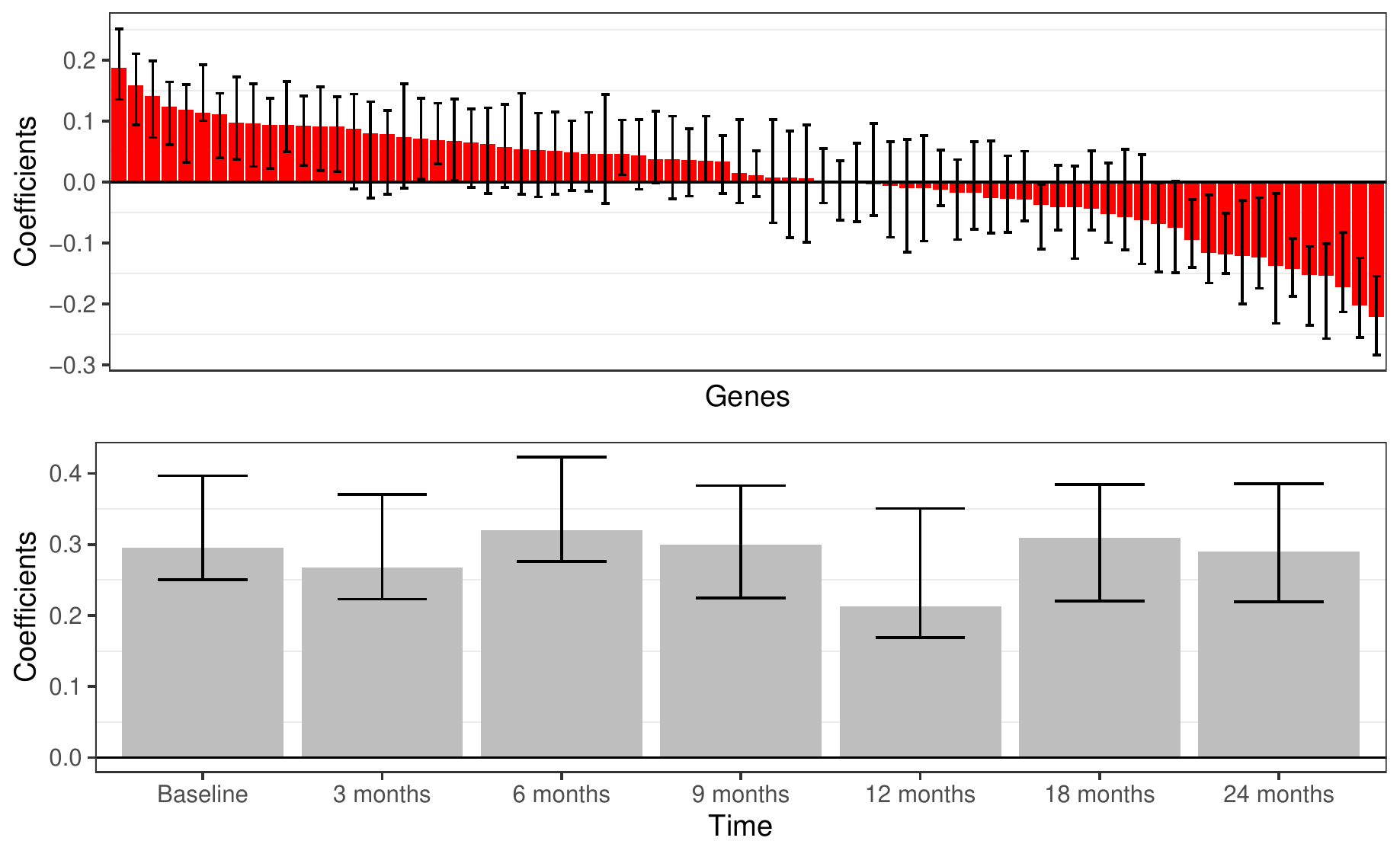}
    \caption{MS treatment study: Rank-1 multiway sparse DWD weights for genes and time with $95\%$ bootstrap confidence intervals. }
    \label{genefig2}
\end{figure}

We further applied multiway sparse DWD to the gene expression time course data described in \cite{baranzini2004transcription}. The purpose of this study is to classify clinical response to treatment for multiple sclerosis (MS) patients. Recombinant human interferon beta (rIFN$\beta$) was given to 53 patients for controlling the symptoms of MS. For each patient, gene expression was measured for 76 genes at 7 time points: baseline (i.e. before treatment) and 6 follow-up time points (3 months, 6 months, 9 months, 12 months, 18 months, and 24 months). The data are a multi-way array with 3 dimensions: patients $\times$ genes $\times$ time points, $\XX: 53 \times 76 \times 7$. Based on clinical characteristics, each patient was designated as a good responder or poor responder to rIFN$\beta$. The proposed rank-1 multiway sparse DWD model was used to differentiate good responders from poor responders. Figure \ref{genefig1} shows the DWD scores under leave-one-out cross validation for the rank-1 multiway model. The groups of these good and poor responders are nearly perfectly separated. The coefficient estimates and 95$\%$ bootstrap confidence intervals for each gene and each time point are shown in Figure \ref{genefig2}. The coefficients across time had little variability, which suggests that the distinction between good and poor responders is not driven by effects that vary over the time course. This agrees with the results reported in \cite{baranzini2004transcription} where they found there was no group*time interaction effects. Note that the data are not very sparse as almost all estimated coefficients for gene and time are not zero. We also compared the performance of the rank-1 model with other competitors. The mean misclassification rates were $13.0\%$ for M-SDWD, $12.6\%$ for M-SDWD with $\lambda_1=0$, $17.4\%$ for M-DWD, $20.8\%$ for CATCH, $20.5\%$ for Full SDWD and $31.5\%$ for Random Forest. The rank-1 multiway sparse models have the best classification over other methods. It is notable that the multiway sparse DWD model outperforms multiway DWD even when the degree of sparsity in the data is low. We also considered the rank-2 multiway sparse model to classify good and poor responders. A rank-2 model performs worse than the rank-1 model with a higher misclassification rate $(18.7\%)$, which agrees with the results in \cite{lyu2017discriminating}. 


\section{Discussion}
\label{sec:discussion}

We have proposed a general framework for high dimensional classification on multiway data with any number of dimensions, which can account for sparsity in the model. Both the simulation and data analysis results have shown that the proposed multiway sparse DWD model can improve classification accuracy when the underlying signal is sparse or has multiway structure.  The proposed method is robust to any degree of sparsity in the model, which has been demonstrated in both simulation and applications. Moreover, the method is robust to the complexity of the shared signal across the different dimensions, by allowing for higher rank models. Lastly, the use of multiway structure and sparsity can facilitate and simplify interpretation.  For our motivating application in Section 6.1, a single application of multiway sparse DWD provided the following insights: (1) TG mice show a distinct brain metabolomic profile after receiving doxycycline but WT mice do not, (2) these changes to the metabolite profile are similar across three neurological regions but most pronounced in the cerebellum, and (3) these changes are most prominent at the end of 12 weeks of treatment, but subside post-treatment.  

Despite the flexibility of multiway sparse DWD, the results are sensitive to the choice of rank and tuning parameters. The simulation in Section~\ref{sim:rank1} shows that when the model is not sparse, the rank-1 model with $\lambda_1=0$ performs better than the rank-1 model with $\lambda_1$ selected by cross validation; thus, the sparsity penalty only improves results when the signal is truly sparse.  Also, the rank-1 model performs poorly when the true model has higher rank, as expected.  Thus, in practice we suggest applying different versions of the method and comparing their performance for any given data application.  Also, as discussed in Sections 4.3, 5.2, and Appendix~\ref{app_convergence}, algorithms using cyclic coordinate descent, including ours, may not converge to a global optimum.  In practice, we suggest running the algorithm from multiple starting values to assess the stability of the results and (if results are sensitive to starting values) select the solution that yields the smallest value for the objective function. In this article we only consider binary classification, and extensions to multi-category classification (e.g., as in \citet{huang2013multiclass}) is a worthwhile future direction.

\section*{Acknowledgements}
This work was supported in part by the National Institutes of Health (NIH) grant R01GM130622. The study on Friedreich's Ataxia mice was supported by the Friedreich's Ataxia Research Alliance (FARA). 

\newpage
\appendix

\section*{Appendices:}

\section{Evaluation of cross-validation method for selecting tuning parameters}
\label{app_cv}

We conducted a simulation study to assess how the choices of tuning parameters $\lambda_1$ and $\lambda_2$ affect the performance of the proposed method. A dataset with sample size $N=100$ was generated, with two classes of equal size $(N_0 = N_1 = 50)$. The predictors have the form of a three-way array of dimensions $\XX: P_1\times P_2\times P_3$ where $P_1=4, P_2=5,$ and $P_3=15$. The $N_0$ samples corresponding to class -1 were generated from a multivariate normal distribution N($\mathbf{0}$,$\mathbf{I}_{P_1P_2P_3\times P_1P_2P_3}$). The other $N_1$ samples corresponding to class 1 were generated from a multivariate normal distribution N($\boldsymbol \mu_1$,$\mathbf{I}_{P_1P_2P_3\times P_1P_2P_3}$) where $\boldsymbol \mu_1 = \mathbf{u}_1 \circ \mathbf{u}_2 \circ \mathbf{u}_3$. Here $\mathbf{u}_1$ and $\mathbf{u}_2$ were generated from $N(\mathbf{0}, \mathbf{I}_{P_k\times P_k})$, $k=1,2$. For $\mathbf{u}_3$ 5 values are set to zero and 10 nonzero values are generated from $N(\mathbf{0}, \mathbf{I}_{10\times 10})$. We consider 4 candidates for $\lambda_1$ $(10^{-4}, 0.001, 0.005, 0.01)$ and 3 candidates for $\lambda_2$ $(1, 0.5, 0.1)$. We applied the multiway sparse DWD method to the simulated data with fixed parameters and selected parameters by cross-validation. The simulation is repeated 100 times. The results are shown in Table \ref{tab_sim_lambda} and Table \ref{tab_sim_cv} with average correlations between true values and estimates for $\mathbf{u}_k$ and average percentages of zero or non-zero coefficients in $\mathbf{u}_3$ that are correctly estimated. The accuracy of classification decreases as $\lambda_2$ decreases. For fixed $\lambda_2$, as $\lambda_1$ becomes larger, more zero coefficients for $\mathbf{u}_3$ are correctly shrunk to 0. Table \ref{tab_sim_cv} shows simulations where the correlations between true values and estimates are very high, and the classification is accurate using the cross-validation method to select parameters, although the selected parameters might be slightly different for each replicate.

\begin{table}[!ht]
\centering
\caption{Simulation results based on multiway sparse DWD with prespecified $\lambda_1$ and $\lambda_2$ candidates. ``Cor($\mathbf{u}_k$)'' is the correlation between the estimated linear hyperplane and the true hyperplane for $k^{th}$ dimension. ``TP($\mathbf{u_3}$)'' is the true positive rate that is the proportion of non-zero coefficients in $\mathbf{u}_3$ are correctly estimated to be non-zero.``TN($\mathbf{u}_3$)'' is the true negative rate that is the proportion of zero coefficients in $\mathbf{u}_3$ are correctly estimated to be zero. The results are the mean over 100 replicates. }
\scalebox{0.8}{
\begin{tabular}{lrrrr|rrrr|rrrr}
  \hline
 $\lambda_2$ &\multicolumn{4}{c|}{1} & \multicolumn{4}{c|}{0.5}& \multicolumn{4}{c}{0.1}\\
$\lambda_1$ & 1e-4 & 0.001 & 0.005 & 0.01 & 1e-4 & 0.001 & 0.005 & 0.01 & 1e-4 & 0.001 & 0.005 & 0.01\\ 
  \hline
Cor($\mathbf{u}_1$) & 0.96 & 0.96 & 0.97 & 0.97 & 0.93 & 0.93 & 0.93 & 0.93 & 0.89 & 0.91 & 0.93 & 0.92 \\ 
  Cor($\mathbf{u}_2$) & 1.00 & 1.00 & 1.00 & 0.99 & 1.00 & 0.99 & 0.99 & 0.98 & 0.95 & 0.95 & 0.95 & 0.93 \\ 
  Cor($\mathbf{u}_3$) & 0.98 & 0.99 & 0.98 & 0.98 & 0.97 & 0.97 & 0.97 & 0.96 & 0.94 & 0.93 & 0.93 & 0.92 \\ 
  TP ($\mathbf{u}_3$)& 1.00 & 0.98 & 0.81 & 0.85 & 1.00 & 0.98 & 0.90 & 0.83  & 0.99 & 0.94 & 0.80 & 0.72\\ 
  TN ($\mathbf{u}_3$) & 0.03 & 0.37 & 0.81 & 0.91 & 0.03 & 0.41 & 0.81 & 0.92& 0.09 & 0.59 & 0.90 & 0.95 \\ 
   \hline
\end{tabular}}
\label{tab_sim_lambda}
\end{table}

\begin{table}[!ht]
\centering
\caption{Simulation results based on multiway sparse DWD with selected penalty parameters by cross-validation. $\lambda_1^G$ and $\lambda_2^G$ denote the geometric means of selected parameters over 100 replicates.}
\scalebox{0.8}{
\begin{tabular}{lc}
  \hline
  & $\lambda_1^G=0.0008, \lambda_2^G=1.56$ \\ 
  \hline
Cor($\mathbf{u}_1$) & 0.98 \\ 
Cor($\mathbf{u}_2$) & 0.99 \\ 
Cor($\mathbf{u}_3$) & 0.99 \\
  TP($\mathbf{u}_3$) & 0.39 \\ 
  TN($\mathbf{u}_3$)  & 0.93 \\ 
   \hline
\end{tabular}}
\label{tab_sim_cv}
\end{table}

\section{Simulation results based on the objective function with separable $L_2$ penalty}
\label{sep_l2_sims}

\begin{figure}
    \centering
    \includegraphics[scale=0.7]{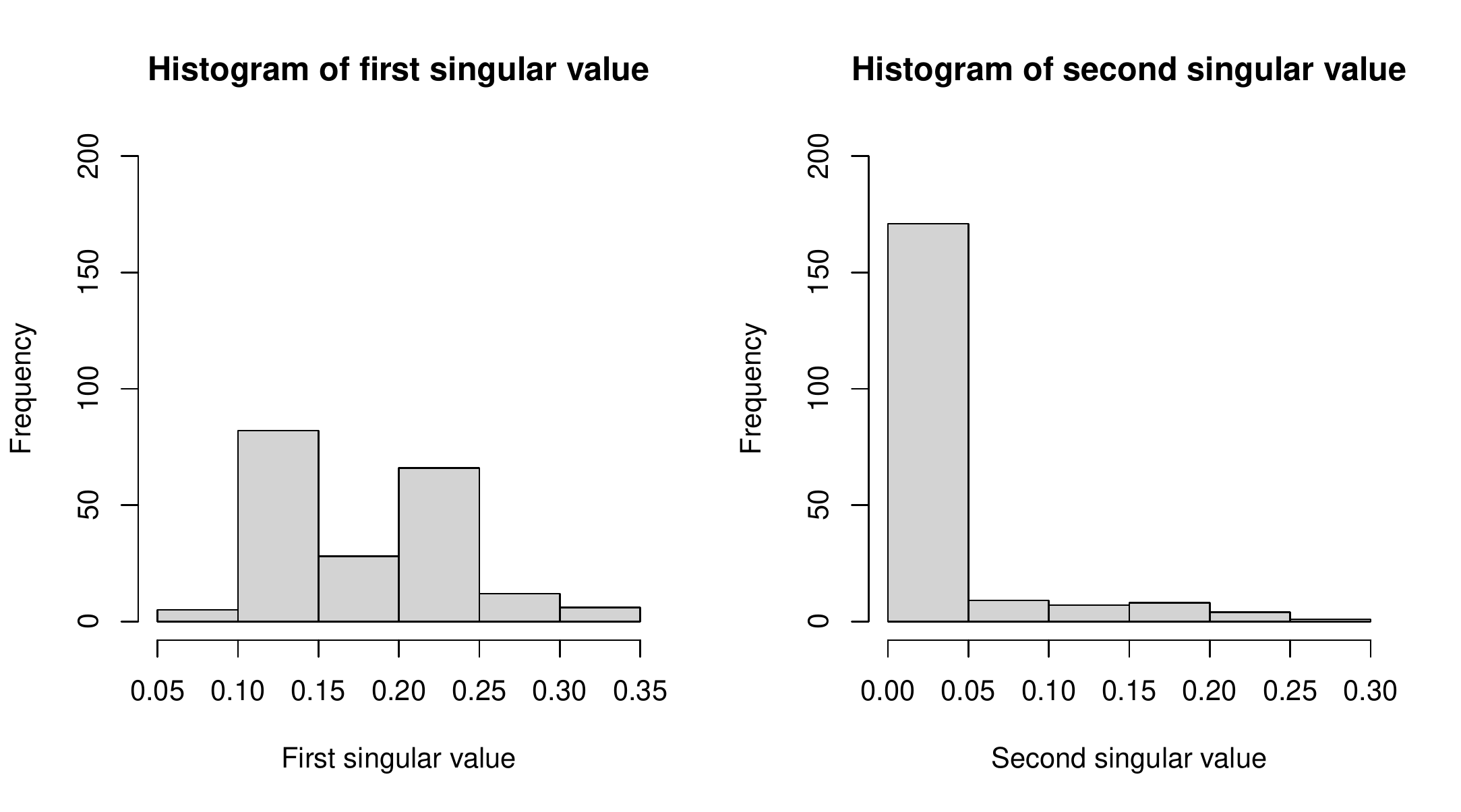}
    \caption{Histograms of the two singular values computed based on the estimated coefficient matrix.}
    \label{fig_svd}
\end{figure}

In this section, we consider a simulation study to evaluate the performance of the higher rank model based on the objective function with separable $L_2$ penalty (i.e. $P^U_{\lambda_{1}, \lambda_{2}}(\mathbb{B})$ defined in the main article). We observed that the separable $L_2$ penalty tends to reduce the rank of the estimated coefficient array, motivating us to use the non-separable $L_2$ penalty (i.e. $P_{\lambda_{1}, \lambda_{2}}(\mathbb{B})$) for the higher rank model. We simulated 200 datasets with high dimensions ($30\times 15 \times 15$) under the case of no sparsity. The process of generating multiway data is identical to that for the higher rank model ($R=2$) described in Section 5.3 of the main article. Table \ref{tab_svd1} shows average correlations, misclassification rates and true positive rates over 200 simulation replicates. Multiway sparse DWD (M-SDWD) with $R=2$ has similar performance to M-SDWD with R=1, which implies M-SDWD (R=2) may not recover the true rank specified in the model. To measure the rank of the estimated model, we conduct SVD to obtain singular values for the estimated coefficient matrix. In most simulations, the second singular values of the estimated coefficient matrix are zero or nearly zero, as shown in Figure \ref{fig_svd}, which indicates the solutions of higher rank (R=2) model are often shrunk to a lower rank. The proportion of simulations that give estimates with true rank $R=2$ is only $14.5 \%$. Table \ref{tab_svd2} shows the results of simulations that can truly detect rank 2 components, and we can see the model with R=2 performs much better than rank-1 multiway models and full SDWD.

\begin{table}
\centering
\caption{Simulation results based on all simulations under the high dimensional scenario: ``Cor'' is the correlation between the estimated linear hyperplane and the true hyperplane. ``Mis'' is the average misclassification rate. ``TP'' is the true positive rate, i.e., the proportion of non-zero coefficients that are correctly estimated to be non-zero.The margins of error (2* standard deviations across 200 replicates) for each statistic are also listed following the $\pm$.}

\begin{tabular}{lrrr}
  \hline
Methods & Cor & Mis & TP  \\ 
  \hline
M-SDWD (R=2) & 0.833$\pm$0.012 & 0.000$\pm$0.000 & 0.883$\pm$0.029 \\ 
  M-SDWD (R=1) & 0.792$\pm$0.011 & 0.000$\pm$0.000 & 0.872$\pm$0.029 \\ 
  M-SDWD ($\lambda_1=0$, R=1) & 0.800$\pm$0.010 & 0.000$\pm$0.000 & 1.000$\pm$0.000 \\ 
  M-DWD & 0.771$\pm$0.014 & 0.000$\pm$0.000 & 1.000$\pm$0.000\\ 
  Full SDWD & 0.781$\pm$0.006 & 0.000$\pm$0.000 & 0.419$\pm$0.036 \\ 
   \hline
\end{tabular}
\label{tab_svd1}
\end{table}

\begin{table}[ht]
\centering
\caption{Simulation results among simulations that give estimates with true rank (R=2) under the high-dimensional scenario: ``Cor'' is the correlation between the estimated linear hyperplane and the true hyperplane. ``Mis'' is the average misclassification rate. ``TP'' is the true positive rate, i.e., the proportion of non-zero coefficients that are correctly estimated to be non-zero.``TN'' is the true negative rate, i.e., the proportion of zero variables that are correctly estimated to be zero. The margins of error (2* standard deviations across 200 replicates) for each statistic are also listed following the $\pm$ symbol.} 

\begin{tabular}{lrrrrr}
  \hline
Methods & Cor & Mis & TP & $\%$ rank 2 \\ 
  \hline
M-SDWD (R=2) & 0.984$\pm$0.008 & 0.000$\pm$0.000 & 0.955$\pm$0.040  & 0.145 \\ 
  M-SDWD (R=1) & 0.709$\pm$0.033 & 0.000$\pm$0.000 & 0.844$\pm$0.088  & - \\ 
  M-SDWD ($\lambda_1=0$, R=1) & 0.726$\pm$0.023 & 0.000$\pm$0.000 & 1.000$\pm$0.000  & - \\ 
  M-DWD & 0.676$\pm$0.024 & 0.000$\pm$0.000 & 1.000$\pm$0.000 & - \\ 
  Full SDWD & 0.782$\pm$0.016 & 0.000$\pm$0.000 & 0.492$\pm$0.088  & - \\ 
   \hline
\end{tabular}
\label{tab_svd2}
\end{table}

\section{Simulation results under different signal to noise ratios}
\label{app_s2n}

We conducted more simulation studies to compare the proposed method with other methods under different signal to noise ratios (SNR= (0.1,0.2,0.3,0.5)). We considered the high-dimensional, more sparsity case of Section 5.1 in the main article for this simulation. Tables \ref{tab_high_snr} and \ref{tab_high_snr_5} show the results based on all simulations and those simulations with correlations larger than 0.5, respectively. The multiway sparse DWD model performs better than other methods in terms of its correlation with the true hyperplane, and has competive misclassification rates across the different SNR levels. The last column in Table \ref{tab_high_snr_5} show the proportions of simulations with correlations larger than 0.5 for different methods under different SNRs. As the SNR increases, the proportion increases as well, as the algorithm tends to converge to the true solution with more signal.

\begin{table}[!ht]
\centering
\caption{Simulation results for different signal to noise ratios: ``Cor'' is the correlation between the estimated linear hyperplane and the true hyperplane. ``Mis'' is the average misclassification rate. ``TP'' is the true positive rate, i.e., the proportion of non-zero coefficients that are correctly estimated to be non-zero.``TN'' is the true negative rate, i.e., the proportion of zero variables that are correctly estimated to be zero. The margins of error (2* standard deviations across 200 replicates) for each statistic are also listed following the $\pm$ symbol.}

\begin{tabular}{llcccc}
  \hline
SNR & Methods & Cor & Mis & TP & TN \\ 
  \hline
0.1 & M-SDWD & 0.717$\pm$0.053 & 0.225$\pm$0.029 & 0.484$\pm$0.050 & 0.837$\pm$0.038 \\ 
    & M-SDWD ($\lambda_1=0$) & 0.594$\pm$0.054 & 0.220$\pm$0.028 & 1.000$\pm$0.000 & 0.000$\pm$0.000 \\ 
    & M-DWD & 0.560$\pm$0.058 & 0.235$\pm$0.030 & 1.000$\pm$0.000 & 0.000$\pm$0.000 \\ 
    & Full SDWD & 0.469$\pm$0.040 & 0.251$\pm$0.025 & 0.168$\pm$0.028 & 0.917$\pm$0.029 \\ 
  0.2 & M-SDWD & 0.849$\pm$0.038 & 0.089$\pm$0.020 & 0.668$\pm$0.040 & 0.806$\pm$0.041 \\ 
    & M-SDWD ($\lambda_1=0$) & 0.796$\pm$0.040 & 0.101$\pm$0.021 & 1.000$\pm$0.000 & 0.000$\pm$0.000 \\ 
    & M-DWD & 0.766$\pm$0.049 & 0.121$\pm$0.025 & 1.000$\pm$0.000 & 0.000$\pm$0.000 \\ 
    & Full SDWD & 0.636$\pm$0.035 & 0.136$\pm$0.022 & 0.172$\pm$0.023 & 0.957$\pm$0.022 \\ 
  0.3 & M-SDWD & 0.879$\pm$0.035 & 0.066$\pm$0.018 & 0.720$\pm$0.036 & 0.776$\pm$0.045 \\ 
    & M-SDWD ($\lambda_1=0$) & 0.838$\pm$0.037 & 0.077$\pm$0.020 & 1.000$\pm$0.000 & 0.000$\pm$0.000 \\ 
    & M-DWD & 0.831$\pm$0.042 & 0.084$\pm$0.021 & 1.000$\pm$0.000 & 0.000$\pm$0.000 \\ 
    & Full SDWD & 0.690$\pm$0.031 & 0.104$\pm$0.020 & 0.165$\pm$0.017 & 0.982$\pm$0.011 \\ 
  0.5 & M-SDWD & 0.927$\pm$0.026 & 0.039$\pm$0.014 & 0.777$\pm$0.033 & 0.762$\pm$0.048 \\ 
    & M-SDWD ($\lambda_1=0$) & 0.890$\pm$0.030 & 0.048$\pm$0.016 & 1.000$\pm$0.000 & 0.000$\pm$0.000 \\ 
    & M-DWD & 0.884$\pm$0.036 & 0.053$\pm$0.018 & 1.000$\pm$0.000 & 0.000$\pm$0.000 \\ 
    & Full SDWD & 0.760$\pm$0.027 & 0.066$\pm$0.017 & 0.189$\pm$0.017 & 0.993$\pm$0.002 \\ 
   \hline
\end{tabular}
\label{tab_high_snr}
\end{table}

\begin{table}[!ht]
\centering
\caption{Simulation results among simulations with correlation greater than 0.5 under different signal to noise ratios: ``Cor'' is the correlation between the estimated linear hyperplane and the true hyperplane. ``Mis'' is the average misclassification rate. ``TP'' is the true positive rate, i.e., the proportion of non-zero coefficients that are correctly estimated to be non-zero.``TN'' is the true negative rate, i.e., the proportion of zero variables that are correctly estimated to be zero. The margins of error (2* standard deviations across 200 replicates) for each statistic are also listed following the $\pm$ symbol.}
\scalebox{0.9}{
\begin{tabular}{lllcccc}
  \hline
SNR & Methods & Cor & Mis & TP & TN & \% Cor$>$0.5 \\ 
  \hline
0.1 & M-SDWD & 0.916$\pm$0.014 & 0.077$\pm$0.015 & 0.628$\pm$0.046 & 0.869$\pm$0.040 & 0.645 \\ 
    & M-SDWD ($\lambda_1=0$) & 0.858$\pm$0.017 & 0.088$\pm$0.016 & 1.000$\pm$0.000 & 0.000$\pm$0.000 & 0.66 \\ 
    & M-DWD & 0.878$\pm$0.018 & 0.080$\pm$0.016 & 1.000$\pm$0.000 & 0.000$\pm$0.000 & 0.62 \\ 
    & Full SDWD & 0.718$\pm$0.023 & 0.096$\pm$0.017 & 0.124$\pm$0.014 & 0.993$\pm$0.002 & 0.52 \\ 
  0.2 & M-SDWD & 0.935$\pm$0.010 & 0.048$\pm$0.010 & 0.697$\pm$0.038 & 0.831$\pm$0.042 & 0.905 \\ 
    & M-SDWD ($\lambda_1=0$) & 0.899$\pm$0.012 & 0.051$\pm$0.010 & 1.000$\pm$0.000 & 0.000$\pm$0.000 & 0.88 \\ 
    & M-DWD & 0.926$\pm$0.010 & 0.042$\pm$0.009 & 1.000$\pm$0.000 & 0.000$\pm$0.000 & 0.825 \\ 
    & Full SDWD & 0.760$\pm$0.017 & 0.060$\pm$0.011 & 0.156$\pm$0.016 & 0.994$\pm$0.001 & 0.765 \\ 
  0.3 & M-SDWD & 0.948$\pm$0.008 & 0.033$\pm$0.008 & 0.745$\pm$0.034 & 0.795$\pm$0.046 & 0.925 \\ 
    & M-SDWD ($\lambda_1=0$) & 0.921$\pm$0.010 & 0.035$\pm$0.009 & 1.000$\pm$0.000 & 0.000$\pm$0.000 & 0.905 \\ 
    & M-DWD & 0.941$\pm$0.008 & 0.030$\pm$0.008 & 1.000$\pm$0.000 & 0.000$\pm$0.000 & 0.88 \\ 
    & Full SDWD & 0.781$\pm$0.016 & 0.048$\pm$0.010 & 0.170$\pm$0.016 & 0.994$\pm$0.002 & 0.825 \\ 
  0.5 & M-SDWD & 0.963$\pm$0.006 & 0.021$\pm$0.007 & 0.792$\pm$0.030 & 0.767$\pm$0.049 & 0.96 \\ 
    & M-SDWD ($\lambda_1=0$) & 0.943$\pm$0.009 & 0.022$\pm$0.007 & 1.000$\pm$0.000 & 0.000$\pm$0.000 & 0.94 \\ 
    & M-DWD & 0.958$\pm$0.007 & 0.018$\pm$0.006 & 1.000$\pm$0.000 & 0.000$\pm$0.000 & 0.92 \\ 
    & Full SDWD & 0.816$\pm$0.015 & 0.030$\pm$0.009 & 0.202$\pm$0.018 & 0.994$\pm$0.001 & 0.89 \\ 
   \hline
\end{tabular}}
\label{tab_high_snr_5}
\end{table}

\section{Assessment of convergence}
\label{app_convergence}

\begin{figure}[!ht]
    \centering
    \includegraphics[scale =0.9]{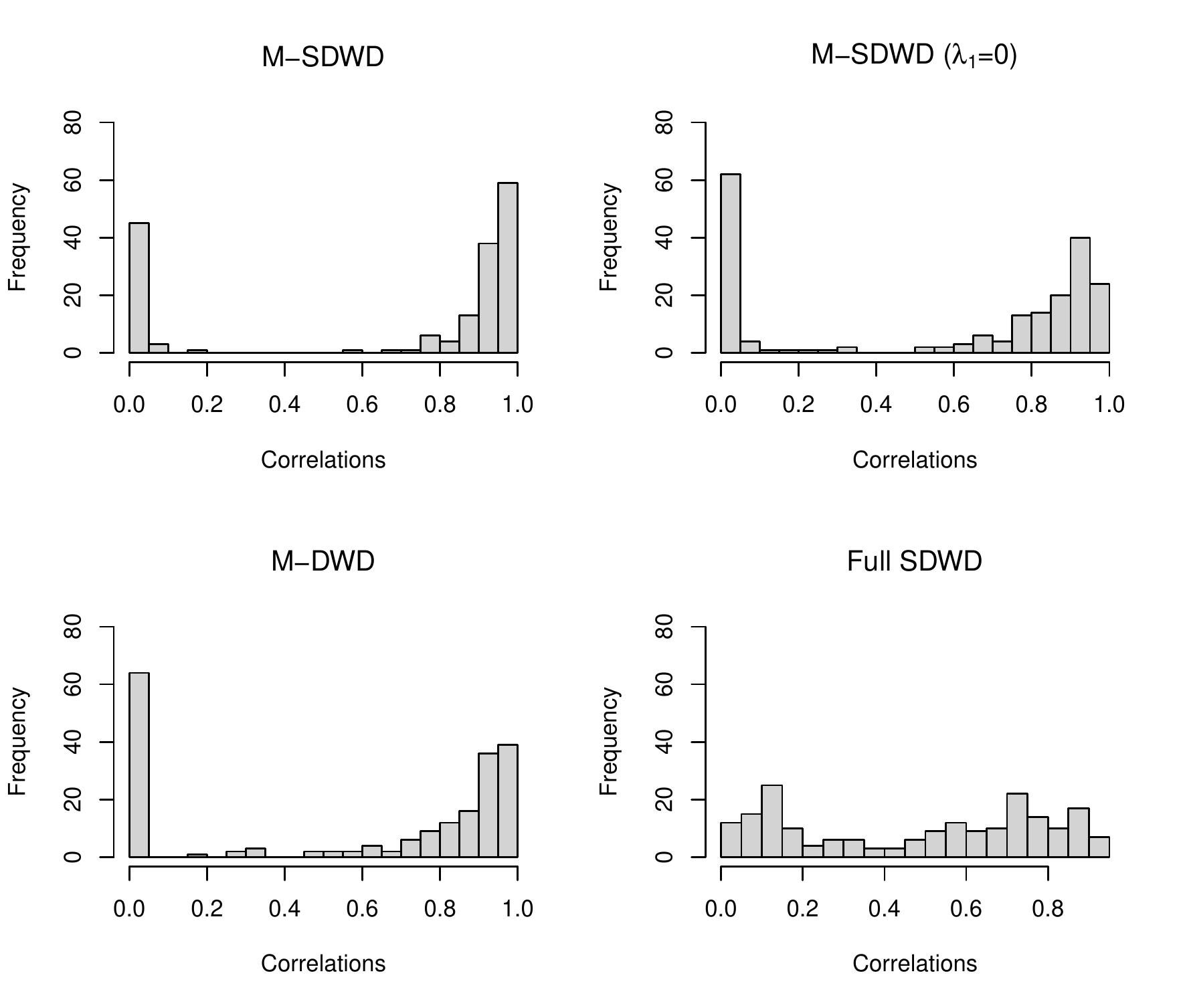}
    \caption{Histogram of correlations between true hyperplane and estimates with four classification methods under high-dimensional scenario with SNR=0.1 and sample size N= 100.}
    \label{fig_hist_1}
\end{figure}

\begin{figure}[!ht]
    \centering
    \includegraphics[scale =0.8]{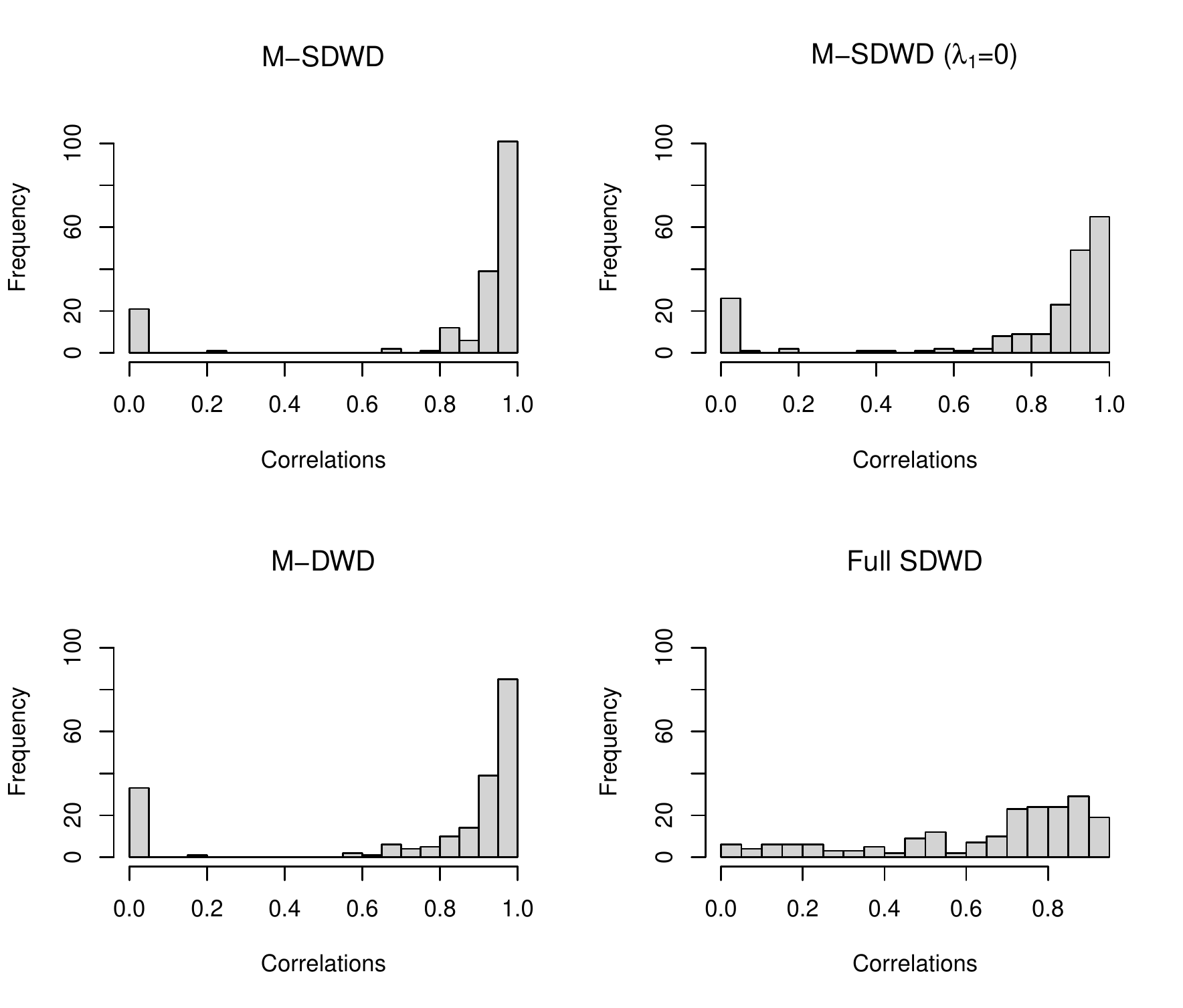}
    \caption{Histogram of correlations between true hyperplane and estimates with four classification methods under high-dimensional scenario with SNR=0.3 and sample size N= 100.}
    \label{fig_hist_3}
\end{figure}

We conducted more simulations to explore how the signal to noise ratio (SNR) affects the convergence of the proposed methods and other alternatives. Figure \ref{fig_hist_1} and \ref{fig_hist_3} show the distributions of correlations between the true hyperplane and the estimated hyperplane for SNR = 0.1 and SNR=0.3, under the more sparsity case based on high dimensional data ($30\times 15 \times 15$) with sample size $N=100$. Combining with results for SNR=0.2 shown in the main article we can see as the SNR increases, the number of correlations with small values decreases, and the convergence issue becomes less severe.

\section{Additional simulations with Rank $>$ 1}
\label{app_highrank}

We conducted more simulations to evaluate the performances of the rank-$R$ multiway sparse model.  These simulations were analogous to those in Section 5.3 of the main article,  with different manipulated conditions.  
Table \ref{tabrank2} gives the results under a lower dimensional setting ($\XX: 15 \times 4 \times 5$).  Table \ref{tab_rank5} gives results for both the lower-dimensional and higher-dimensional settings for a model with rank $R=5$.  From these results we can conclude that the rank-$R$ model can generally perform well when the data were generated under a true rank-$R$ model, although lower-rank approximations perform comparably well when the dimension is small and the signal is sparse. 

Table \ref{tab_diffranks} shows the mean correlations between the estimated linear hyperplane and the true hyperplane using the proposed method (M-SDWD) by assumed rank ($\hat{R} = 1, 2, 3, 4, 5$) for different true ranks under the high dimensional scenario with more sparsity when N = 40. In general, the correlation is maximized when the assumed rank is equal to the true rank; however, performance tends to be robust to deviations from the precise value of the true rank, especially when the true rank is higher.   

\begin{table}[!ht]
\centering
\caption{Simulation results under the low dimensional scenario ($15 \times 4 \times 5$) when the true model is rank-2. In the Sparsity column, the numbers in  parentheses indicate the number of non-zero variables in each dimension. ``Cor'' is the correlation between the estimated linear hyperplane and the true hyperplane. ``Mis'' is the average misclassification rate. ``TP'' is the true positive rate, i.e., the proportion of non-zero coefficients that are correctly estimated to be non-zero.``TN'' is the true negative rate, i.e., the proportion of zero coefficients that are correctly estimated to be zero. The margins of error (2* standard deviations across 200 replicates) for each statistic are also listed following the $\pm$ symbol.}
\scalebox{0.8}{
\begin{tabular}{lllllll}
  \hline
N & Sparsity & Methods & Cor & Mis & TP & TN \\ 
  \hline
40& More ($5 \times 2 \times 2$)  & M-SDWD (R=2) & 0.716$\pm$0.034 & 0.191$\pm$0.027 & 0.659$\pm$0.052 & 0.678$\pm$0.048 \\     &   & M-SDWD ($\lambda_1=0$, R=2 ) & 0.643$\pm$0.032 & 0.183$\pm$0.023 & 1.000$\pm$0.000 & 0.000$\pm$0.000 \\ 
    &   & M-SDWD (R=1) & {\bf 0.746$\pm$0.038} & {\bf 0.157$\pm$0.024} & 0.676$\pm$0.052 & 0.711$\pm$0.056 \\ 
    &   & M-SDWD ($\lambda_1=0$, R=1 ) & 0.713$\pm$0.036 & 0.167$\pm$0.024 & 1.000$\pm$0.000 & 0.000$\pm$0.000 \\ 
    &   & M-DWD & 0.724$\pm$0.041 & 0.168$\pm$0.025 & 1.000$\pm$0.000 & 0.000$\pm$0.000 \\ 
    &   & Full SDWD & 0.618$\pm$0.033 & 0.188$\pm$0.023 & 0.410$\pm$0.034 & 0.861$\pm$0.033 \\ 
    & Less ($5 \times 4 \times 5$)   & M-SDWD (R=2) & {\bf 0.875$\pm$0.014} & 0.038$\pm$0.010 & 0.786$\pm$0.036 & 0.440$\pm$0.052 \\ 
    &   & M-SDWD ($\lambda_1=0$, R=2 ) & 0.870$\pm$0.015 & {\bf 0.036$\pm$0.009} & 1.000$\pm$0.000 & 0.000$\pm$0.000 \\ 
    &   & M-SDWD (R=1) & 0.810$\pm$0.017 & 0.043$\pm$0.010 & 0.698$\pm$0.042 & 0.530$\pm$0.054 \\ 
    &   & M-SDWD ($\lambda_1=0$, R=1 ) & 0.818$\pm$0.015 & 0.040$\pm$0.009 & 1.000$\pm$0.000 & 0.000$\pm$0.000 \\ 
    &   & M-DWD & 0.824$\pm$0.017 & 0.039$\pm$0.010 & 1.000$\pm$0.000 & 0.000$\pm$0.000 \\ 
    &   & Full SDWD & 0.703$\pm$0.019 & 0.064$\pm$0.014 & 0.334$\pm$0.032 & 0.847$\pm$0.033 \\ 

    & No ($15 \times 4 \times 5$) & M-SDWD (R=2) & {\bf 0.958$\pm$0.007} & 0.003$\pm$0.003 & 1.000$\pm$0.000 &  - \\ 
    &   & M-SDWD ($\lambda_1=0$, R=2 ) & {\bf 0.959$\pm$0.006} & 0.003$\pm$0.002 & 1.000$\pm$0.000 &  - \\ 
    &   & M-SDWD (R=1) & 0.840$\pm$0.014 & 0.006$\pm$0.003 & 0.839$\pm$0.034 &  - \\ 
    &   & M-SDWD ($\lambda_1=0$, R=1 ) & 0.849$\pm$0.012 & 0.005$\pm$0.003 & 1.000$\pm$0.000 &  - \\ 
    &   & M-DWD & 0.843$\pm$0.014 & 0.005$\pm$0.003 & 1.000$\pm$0.000 &  - \\ 
    &   & Full SDWD & 0.773$\pm$0.012 & 0.011$\pm$0.005 & 0.469$\pm$0.038 &  - \\ 
100  & More ($5 \times 2 \times 2$)   & M-SDWD (R=2) & 0.853$\pm$0.021 & 0.144$\pm$0.023 & 0.764$\pm$0.046 & 0.702$\pm$0.051 \\ 
    &   & M-SDWD ($\lambda_1=0$, R=2 ) & 0.791$\pm$0.026 & 0.156$\pm$0.023 & 1.000$\pm$0.000 & 0.000$\pm$0.000 \\ 
    &   & M-SDWD (R=1) & {\bf 0.862$\pm$0.022} & {\bf 0.141$\pm$0.020} & 0.735$\pm$0.052 & 0.763$\pm$0.059 \\ 
    &   & M-SDWD ($\lambda_1=0$, R=1 ) & 0.840$\pm$0.022 & 0.148$\pm$0.021 & 1.000$\pm$0.000 & 0.000$\pm$0.000 \\ 
    &   & M-DWD & 0.834$\pm$0.031 & 0.148$\pm$0.022 & 1.000$\pm$0.000 & 0.000$\pm$0.000 \\ 
    &   & Full SDWD & 0.798$\pm$0.027 & 0.160$\pm$0.025 & 0.395$\pm$0.037 & 0.934$\pm$0.025 \\ 
 
    & Less ($5 \times 4 \times 5$)   & M-SDWD (R=2) & {\bf 0.941$\pm$0.009} & {\bf 0.027$\pm$0.007} & 0.858$\pm$0.029 & 0.410$\pm$0.050 \\ 
    &   & M-SDWD ($\lambda_1=0$, R=2 ) & 0.938$\pm$0.009 & 0.027$\pm$0.007 & 1.000$\pm$0.000 & 0.000$\pm$0.000 \\ 
    &   & M-SDWD (R=1) & 0.860$\pm$0.012 & 0.030$\pm$0.007 & 0.772$\pm$0.033 & 0.518$\pm$0.053 \\ 
    &   & M-SDWD ($\lambda_1=0$, R=1 ) & 0.862$\pm$0.012 & 0.031$\pm$0.007 & 1.000$\pm$0.000 & 0.000$\pm$0.000 \\ 
    &   & M-DWD & 0.868$\pm$0.012 & 0.031$\pm$0.007 & 1.000$\pm$0.000 & 0.000$\pm$0.000 \\ 
    &   & Full SDWD & 0.838$\pm$0.011 & 0.037$\pm$0.009 & 0.371$\pm$0.025 & 0.893$\pm$0.019 \\ 
        & No ($15 \times 4 \times 5$)  & M-SDWD (R=2) & 0.976$\pm$0.004 & {\bf 0.004$\pm$0.003} & 0.964$\pm$0.012 &  - \\ 
    &   & M-SDWD ($\lambda_1=0$, R=2 ) & {\bf 0.979$\pm$0.004} & {\bf 0.004$\pm$0.003} & 1.000$\pm$0.000 &  - \\ 
    &   & M-SDWD (R=1) & 0.849$\pm$0.013 & 0.007$\pm$0.003 & 0.880$\pm$0.026 &  - \\ 
    &   & M-SDWD ($\lambda_1=0$, R=1 ) & 0.854$\pm$0.012 & 0.007$\pm$0.003 & 1.000$\pm$0.000 &  - \\ 
    &   & M-DWD & 0.846$\pm$0.015 & 0.007$\pm$0.003 & 1.000$\pm$0.000 &  - \\ 
    &   & Full SDWD & 0.860$\pm$0.009 & 0.009$\pm$0.005 & 0.543$\pm$0.037 &  - \\ 
   \hline
\end{tabular}}
\label{tabrank2}
\end{table}    

    \begin{table}[!ht]
\centering
\caption{Simulation results when the true model is rank-5 (R=5) under high and low dimensional scenarios and sample size $N=40$. In the Sparsity column, the numbers in  parentheses indicate the number of non-zero variables in each dimension. ``Cor'' is the correlation between the estimated linear hyperplane and the true hyperplane. ``Mis'' is the average misclassification rate. ``TP'' is the true positive rate, i.e., the proportion of non-zero coefficients that are correctly estimated to be non-zero.``TN'' is the true negative rate, i.e., the proportion of zero coefficients that are correctly estimated to be zero. The margins of error (2* standard deviations across 200 replicates) for each statistic are also listed following the $\pm$ symbol.}
\label{tab:resR5}
\scalebox{0.7}{
\begin{tabular}{lllllll}
  \hline
Dimensions & Sparsity & Methods & Cor & Mis & TP & TN \\ 
  \hline
High ($30\times 15 \times 15$) & More ($5\times 5 \times 5$) & M-SDWD (R=5) & {\bf 0.845$\pm$0.013} & {\bf 0.002$\pm$0.002} & 0.927$\pm$0.016 & 0.623$\pm$0.048 \\ 
   &   & M-SDWD ($\lambda_1=0$,R=5) & 0.774$\pm$0.017 & 0.007$\pm$0.004 & 1.000$\pm$0.000 & 0.000$\pm$0.000 \\ 
    &   & M-SDWD (R=1) & 0.719$\pm$0.015 & 0.005$\pm$0.003 & 0.718$\pm$0.037 & 0.736$\pm$0.051 \\ 
    &   & M-SDWD ($\lambda_1=0$,R=1) & 0.704$\pm$0.018 & 0.007$\pm$0.006 & 1.000$\pm$0.000 & 0.000$\pm$0.000 \\ 
    &   & M-DWD  & 0.702$\pm$0.020 & 0.010$\pm$0.008 & 1.000$\pm$0.000 & 0.000$\pm$0.000 \\ 
    &   & Full SDWD & 0.732$\pm$0.015 & 0.007$\pm$0.004 & 0.262$\pm$0.016 & 0.993$\pm$0.002 \\ 
    & Less ($10\times 15 \times 15$) & M-SDWD (R=5) & 0.965$\pm$0.006 & 0.000$\pm$0.000 & 0.965$\pm$0.014 & 0.374$\pm$0.062 \\ 
    &   & M-SDWD ($\lambda_1=0$,R=5) & {\bf 0.981$\pm$0.002} & 0.000$\pm$0.000 & 1.000$\pm$0.000 & 0.000$\pm$0.000 \\ 
    &   & M-SDWD (R=1) & 0.598$\pm$0.011 & 0.000$\pm$0.000 & 0.854$\pm$0.031 & 0.450$\pm$0.063 \\ 
    &   & M-SDWD ($\lambda_1=0$,R=1) & 0.603$\pm$0.010 & 0.000$\pm$0.000 & 1.000$\pm$0.000 & 0.000$\pm$0.000 \\ 
    &   & M-DWD  & 0.574$\pm$0.013 & 0.000$\pm$0.000 & 1.000$\pm$0.000 & 0.000$\pm$0.000 \\ 
    &   & Full SDWD & 0.813$\pm$0.005 & 0.000$\pm$0.000 & 0.420$\pm$0.024 & 0.906$\pm$0.020 \\ 
      & No ($30\times 15 \times 15$)  & M-SDWD (R=5) & 0.976$\pm$0.003 & 0.000$\pm$0.000 & 1.000$\pm$0.000 &  - \\ 
    &   & M-SDWD ($\lambda_1=0$,R=5) & {\bf 0.981$\pm$0.003} & 0.000$\pm$0.000 & 1.000$\pm$0.000 &  - \\ 
    &   & M-SDWD (R=1) & 0.573$\pm$0.010 & 0.000$\pm$0.000 & 0.794$\pm$0.037 &  - \\ 
    &   & M-SDWD ($\lambda_1=0$,R=1) & 0.590$\pm$0.008 & 0.000$\pm$0.000 & 1.000$\pm$0.000 &  - \\ 
    &   & M-DWD  & 0.557$\pm$0.012 & 0.000$\pm$0.000 & 1.000$\pm$0.000 &  - \\ 
    &   & Full SDWD & 0.828$\pm$0.006 & 0.000$\pm$0.000 & 0.556$\pm$0.037 &  - \\ 

Low ($15 \times 4 \times 5$)  & More ($5 \times 2 \times 2$)   & M-SDWD (R=5 & 0.733$\pm$0.021 & 0.128$\pm$0.022 & 0.816$\pm$0.042 & 0.472$\pm$0.045 \\ 
    &   & M-SDWD ($\lambda_1=0$,R=5) & 0.664$\pm$0.022 & 0.120$\pm$0.017 & 1.000$\pm$0.000 & 0.000$\pm$0.000 \\ 
    &   & M-SDWD (R=1) & {\bf 0.755$\pm$0.022} & {\bf 0.093$\pm$0.015} & 0.671$\pm$0.047 & 0.703$\pm$0.053 \\ 
    &   & M-SDWD ($\lambda_1=0$,R=1) & 0.731$\pm$0.024 & 0.103$\pm$0.016 & 1.000$\pm$0.000 & 0.000$\pm$0.000 \\ 
    &   & M-DWD  & 0.739$\pm$0.026 & 0.102$\pm$0.017 & 1.000$\pm$0.000 & 0.000$\pm$0.000 \\ 
    &   & Full SDWD & 0.729$\pm$0.023 & 0.096$\pm$0.016 & 0.442$\pm$0.025 & 0.909$\pm$0.023 \\ 
    & Less ($5 \times 4 \times 5$)  & M-SDWD (R=5) & {\bf 0.892$\pm$0.011} & 0.010$\pm$0.007 & 0.922$\pm$0.021 & 0.269$\pm$0.045 \\ 
    &   & M-SDWD ($\lambda_1=0$, R=5) & 0.889$\pm$0.010 & {\bf 0.005$\pm$0.004} & 1.000$\pm$0.000 & 0.000$\pm$0.000 \\ 
    &   & M-SDWD (R=1) & 0.717$\pm$0.015 & 0.010$\pm$0.004 & 0.739$\pm$0.040 & 0.515$\pm$0.057 \\ 
    &   & M-SDWD ($\lambda_1=0$,R=1) & 0.722$\pm$0.014 & 0.011$\pm$0.004 & 1.000$\pm$0.000 & 0.000$\pm$0.000 \\ 
    &   & M-DWD  & 0.717$\pm$0.015 & 0.010$\pm$0.004 & 1.000$\pm$0.000 & 0.000$\pm$0.000 \\ 
    &   & Full SDWD & 0.795$\pm$0.010 & 0.008$\pm$0.004 & 0.468$\pm$0.031 & 0.841$\pm$0.029 \\ 
    & No ($15 \times 4 \times 5$)  & M-SDWD (R=5) & 0.857$\pm$0.009 & 0.000$\pm$0.000 & 0.950$\pm$0.018 &  - \\ 
    &   & M-SDWD ($\lambda_1=0$, R=5) & {\bf 0.864$\pm$0.007} & 0.000$\pm$0.000 & 1.000$\pm$0.000 &  - \\ 
    &   & M-SDWD (R=1) & 0.693$\pm$0.013 & 0.000$\pm$0.000 & 0.836$\pm$0.033 &  - \\ 
    &   & M-SDWD ($\lambda_1=0$,R=1) & 0.701$\pm$0.012 & 0.000$\pm$0.000 & 1.000$\pm$0.000 &  - \\ 
    &   & M-DWD  & 0.687$\pm$0.014 & 0.000$\pm$0.000 & 1.000$\pm$0.000 &  - \\ 
    &   & Full SDWD & 0.841$\pm$0.005 & 0.000$\pm$0.000 & 0.573$\pm$0.035 &  - \\ 
   \hline
\end{tabular}}
\label{tab_rank5}
\end{table}


\begin{table}[!ht]
\centering
\caption{ Mean correlations between the estimated linear hyperplane and the true hyperplane by assumed rank for different true ranks for N=40 under high dimensional scenarios $30 \times 15 \times 15$ with more sparsity ($5 \times 5 \times 5$ of variables has signals have signal discriminating the classes).The assumed rank that return the largest correlation for each true rank is in bold.}
\label{tab_diffranks}
\begin{tabular}{rrrrrr}
  \hline
& $\hat{R}=1$ & $\hat{R}=2$ & $\hat{R}=3$ & $\hat{R}=4$ & $\hat{R}=5$ \\ 
  \hline
$R=1$ & {\bf 0.503} & 0.490 & 0.432 & 0.486 & 0.417 \\ 
  $R=2$ & 0.737 & {\bf 0.803} & 0.766 & 0.779 & 0.726 \\ 
  $R=3$ & 0.731 & 0.804 & {\bf 0.811} & 0.806 & 0.791 \\ 
  $R=4$ & 0.712 & 0.794 & 0.828 & {\bf 0.835} & 0.834 \\ 
  $R=5$ & 0.703 & 0.808 & {\bf0.828} & 0.819 &  0.827 \\ 
   \hline
\end{tabular}
\end{table}


\section{Simulations with correlated predictors}
\label{app_corr}

In this section, we considered a simulation study in which the predictors are correlated.  The covariance matrix for the predictor array is assumed to have a separable structure $\Sigma=\Sigma_1 \otimes \Sigma_2 \otimes \Sigma_3$, where $\Sigma_k: P_k \times P_k$ is the covariance along dimension $k$. Each of $\Sigma_1$, $\Sigma_2$, and $\Sigma_3$ have an AR(1) structure, with correlation determined by a shared parameter $\rho$:
\[\Sigma_k = \begin{bmatrix} 
    1 & \rho & \rho^2 & \dots \\
    & 1 & \rho & \dots & \\
    & & \ddots& & \\
    & & & 1
    \end{bmatrix}
 \; \; \text{ for } k=1,2,3.\]
 Thus, $\rho$ controls the overall level of correlation in the predictors.  The samples for class -1 are generated via $\mbox{vec}(\XX_i)=\text{Normal}(\mathbf{0}, \Sigma)$ and the samples for class $+1$ are generated via $\mbox{vec}(\XX_i)=\text{Normal}(\mbox{vec}(\sqrt{\alpha} \mu_1), \Sigma)$,  with $\sqrt{\alpha} \mu_1$ generated as in Section 5.1 of the main article.  As a representative scenario we consider the high-dimensional ($30 \times 15\times 15$), more sparsity, small sample size (N=40) case of Section 5.1 in the main article, and generate $\sqrt{\alpha} \mu_1$ under those conditions.  Table~\ref{tab_correlated} shows the results over $200$ replications across different methods and different correlation levels $\rho$.  This shows that the M-SDWD method performs relatively well for scenarios with no correlation or mild correlation, but less well for scenarios with high correlation.  We find that the multiway approaches are more prone to convergence to a local optima with high levels of correlation.  Table~\ref{tab_correlatedCor0.5} shows the results for those replications that converge to a reasonable solution (with correlation greater than $0.5$ with the true vector), and the multiway approaches perform better under all levels of residual correlation $\rho$ for these replications.

\begin{table}[!ht]
\centering
\caption{Simulation results under the high dimensional  ($30 \times 15\times 15$), more sparsity, small sample size (N=40) with different residual correlation levels $\rho$. ``Cor'' is the correlation between the estimated linear hyperplane and the true hyperplane. ``Mis'' is the average misclassification rate. ``TP'' is the true positive rate, i.e., the proportion of non-zero coefficients that are correctly estimated to be non-zero.``TN'' is the true negative rate, i.e., the proportion of zero coefficients that are correctly estimated to be zero. The margins of error (2* standard deviations across 200 replicates) for each statistic are also listed following the $\pm$ symbol.}
\label{tab_correlated}
\begin{tabular}{rlllll}
  \hline
$\rho$ & Methods & Cor & Mis & TP & TN \\ 
  \hline
0 & M-SDWD & {\bf 0.562$\pm$0.061} & {\bf 0.204$\pm$0.033} & 0.523$\pm$0.049 & 0.724$\pm$0.051 \\ 
   & M-SDWD ($\lambda_1=0$) & 0.519$\pm$0.058 & 0.205$\pm$0.031 & 1.000$\pm$0.000 & 0.000$\pm$0.000 \\ 
  & M-DWD & 0.516$\pm$0.060 & 0.226$\pm$0.034 & 1.000$\pm$0.000 & 0.000$\pm$0.000 \\ 
  & Full SDWD & 0.347$\pm$0.037 & 0.263$\pm$0.027 & 0.197$\pm$0.033 & 0.884$\pm$0.034 \\ 
  0.3 & M-SDWD & {\bf 0.454$\pm$0.062} & {\bf 0.255$\pm$0.033} & 0.576$\pm$0.052 & 0.649$\pm$0.058 \\ 
   & M-SDWD ($\lambda_1=0$) & 0.436$\pm$0.059 & 0.260$\pm$0.032 & 1.000$\pm$0.000 & 0.000$\pm$0.000 \\ 
   & M-DWD & 0.310$\pm$0.058 & 0.337$\pm$0.032 & 1.000$\pm$0.000 & 0.000$\pm$0.000 \\ 
   & Full SDWD & 0.355$\pm$0.036 & 0.253$\pm$0.026 & 0.193$\pm$0.032 & 0.884$\pm$0.034 \\ 
  0.6 & M-SDWD & 0.323$\pm$0.062 & 0.325$\pm$0.035 & 0.456$\pm$0.057 & 0.680$\pm$0.058 \\ 
   & M-SDWD ($\lambda_1=0$) & 0.292$\pm$0.060 & 0.356$\pm$0.033 & 1.000$\pm$0.000 & 0.000$\pm$0.000 \\ 
   & M-DWD & 0.067$\pm$0.032 & 0.474$\pm$0.019 & 1.000$\pm$0.000 & 0.000$\pm$0.000 \\ 
   & Full SDWD & {\bf 0.375$\pm$0.039} & {\bf 0.267$\pm$0.029} & 0.177$\pm$0.034 & 0.900$\pm$0.035 \\ 
  0.9 & M-SDWD & 0.434$\pm$0.069 & 0.256$\pm$0.040 & 0.393$\pm$0.060 & 0.791$\pm$0.054 \\ 
   & M-SDWD ($\lambda_1=0$) & 0.427$\pm$0.072 & 0.288$\pm$0.040 & 1.000$\pm$0.000 & 0.000$\pm$0.000 \\ 
   & M-DWD & 0.044$\pm$0.027 & 0.477$\pm$0.017 & 1.000$\pm$0.000 & 0.000$\pm$0.000 \\ 
   & Full SDWD & {\bf 0.561$\pm$0.043} & {\bf 0.177$\pm$0.032} & 0.193$\pm$0.040 & 0.897$\pm$0.043 \\ 
   \hline
\end{tabular}
\end{table}

\begin{table}[!ht]
\centering
\caption{Simulation results among replications with ``Cor"$>$0.5 under the high dimensional  ($30 \times 15\times 15$), more sparsity, small sample size (N=40) with different residual correlation levels $\rho$. ``Cor'' is the correlation between the estimated linear hyperplane and the true hyperplane. ``Mis'' is the average misclassification rate. ``TP'' is the true positive rate, i.e., the proportion of non-zero coefficients that are correctly estimated to be non-zero.``TN'' is the true negative rate, i.e., the proportion of zero coefficients that are correctly estimated to be zero. The margins of error (2* standard deviations across 200 replicates) for each statistic are also listed following the $\pm$ symbol.}
\label{tab_correlatedCor0.5}
\scalebox{0.9}{
\begin{tabular}{rlllllr}
  \hline
$\rho$ & Methods & Cor & Mis & TP & TN & $\%$Cor$>0.5$ \\ 
  \hline
 0& M-SDWD & {\bf 0.897$\pm$0.014} & {\bf 0.027$\pm$0.007} & 0.582$\pm$0.050 & 0.832$\pm$0.051 & 0.62 \\ 
   & M-SDWD ($\lambda_1=0$) & 0.841$\pm$0.018 & 0.036$\pm$0.010 & 1.000$\pm$0.000 & 0.000$\pm$0.000 & 0.60 \\ 
   & M-DWD & 0.864$\pm$0.018 & 0.036$\pm$0.010 & 1.000$\pm$0.000 & 0.000$\pm$0.000 & 0.59 \\ 
   & Full SDWD & 0.673$\pm$0.029 & 0.043$\pm$0.015 & 0.131$\pm$0.018 & 0.993$\pm$0.002 & 0.32 \\ 
  0.3 & M-SDWD & {\bf 0.901$\pm$0.016} & {\bf 0.022$\pm$0.008} & 0.601$\pm$0.058 & 0.840$\pm$0.056 & 0.49 \\ 
   & M-SDWD ($\lambda_1=0$) & 0.850$\pm$0.021 & 0.036$\pm$0.011 & 1.000$\pm$0.000 & 0.000$\pm$0.000 & 0.49 \\ 
   & M-DWD & 0.893$\pm$0.022 & 0.028$\pm$0.011 & 1.000$\pm$0.000 & 0.000$\pm$0.000 & 0.33 \\ 
   & Full SDWD & 0.674$\pm$0.028 & 0.042$\pm$0.016 & 0.121$\pm$0.015 & 0.994$\pm$0.001 & 0.32 \\ 
  0.6 & M-SDWD & 0.903$\pm$0.022 & {\bf 0.009$\pm$0.008} & 0.468$\pm$0.079 & 0.912$\pm$0.057 & 0.35 \\ 
   & M-SDWD ($\lambda_1=0$) & 0.906$\pm$0.023 & 0.026$\pm$0.015 & 1.000$\pm$0.000 & 0.000$\pm$0.000 & 0.30 \\ 
   & M-DWD & {\bf 0.956$\pm$0.032} & 0.012$\pm$0.024 & 1.000$\pm$0.000 & 0.000$\pm$0.000 & 0.05 \\ 
   & Full SDWD & 0.679$\pm$0.028 & 0.050$\pm$0.016 & 0.123$\pm$0.018 & 0.995$\pm$0.001 & 0.36 \\ 
 0.9 & M-SDWD & 0.877$\pm$0.026 & {\bf 0.002$\pm$0.002} & 0.475$\pm$0.075 & 0.934$\pm$0.042 & 0.49 \\ 
   & M-SDWD ($\lambda_1=0$) & {\bf 0.935$\pm$0.019} & 0.018$\pm$0.013 & 1.000$\pm$0.000 & 0.000$\pm$0.000 & 0.45 \\ 
   & M-DWD & 0.885$\pm$0.131 & 0.083$\pm$0.110 & 1.000$\pm$0.000 & 0.000$\pm$0.000 & 0.04 \\ 
   & Full SDWD & 0.730$\pm$0.018 & 0.056$\pm$0.016 & 0.117$\pm$0.011 & 0.996$\pm$0.001 & 0.69 \\ 
   \hline
\end{tabular}}
\end{table}

\section{Tuning parameters selection for MRS data application}
\label{app_tuning}

Here we expand on the selection of the tuning parameters for the application of multi-way sparse DWD to classify dox treated mice (dox group) and controls (no dox group) in Section 6.1 of the main article. We computed the predicted DWD scores for all subjects by 10-fold cross-validation for a grid of $\lambda_1$ ($0, 0.0001, 0.001, 0.005, 0.01, 0.025, 0.05)$ and $\lambda_2$($0.25, 0.5, 0.75, 1, 3, 5$), and select the best combinations of parameters with the maximum t-test statistics that testing the differences between the scores and the two classes.
Table \ref{tab_tstatistic} shows the t-test statistics between dox and no-dox groups for a grid of $\lambda_1$ and $\lambda_2$. The best pair of tuning parameters are $\lambda_1=0.01$ and $\lambda_2=5$ with the maximum test statistic. Table \ref{tab_mis} shows the misclassification rates that classify the TG-dox group and other mice. Using t-test statistic as index for selecting parameters gives a unique optimal pair of parameters, which also located in the region of parameters with minimal misclassification rates.




\begin{table}[!ht]
\centering
\centering
\caption{T-test statistics that test the differences of predicted DWD scores between dox and no-dox groups for a grid of $\lambda_1$ and $\lambda_2$}
\begin{tabular}{rrrrrrrrrr}
  \hline
$\lambda_2$/$\lambda_1$ & 0 & 1e-04 & 0.001 & 0.005 & 0.01 & 0.025 & 0.05 & 0.1 & 0.25 \\ 
  \hline
0.25 & 2.098 & 2.105 & 2.194 & 2.537 & 2.789 & 2.759 & 2.823 & 3.189 & 3.549 \\ 
  0.50 & 1.904 & 1.906 & 1.932 & 2.060 & 2.303 & 2.712 & 2.778 & 2.948 & 3.090 \\ 
  0.75 & 1.556 & 1.564 & 1.644 & 1.950 & 2.350 & 2.678 & 2.663 & 2.658 & 3.020 \\ 
  1.00 & 2.676 & 2.677 & 2.684 & 2.694 & 2.762 & 3.032 & 3.024 & 3.069 & 3.455 \\ 
  3.00 & 3.251 & 3.253 & 3.266 & 3.308 & 3.377 & 3.483 & 3.525 & 3.573 & {\bf 3.660} \\ 
  5.00 & 3.172 & 3.172 & 3.180 & 3.216 & 3.264 & 3.327 & 3.384 & 3.511 & 3.427 \\ 
  \hline
\end{tabular}
\label{tab_tstatistic}
\end{table}


\begin{table}[!ht]
\centering
\caption{Misclassification rates that classify TG-dox and other groups for a grid of $\lambda_1$ and $\lambda_2$}
\begin{tabular}{rrrrrrrrrr}
  \hline
$\lambda_2$/$\lambda_1$ & 0 & 1e-04 & 0.001 & 0.005 & 0.01 & 0.025 & 0.05 & 0.1 & 0.25 \\ 
  \hline
0.25 & 0.333 & 0.333 & 0.333 & 0.333 & 0.286 & 0.286 & 0.286 & 0.286 & 0.190 \\ 
  0.50 & 0.286 & 0.286 & 0.286 & 0.286 & 0.286 & 0.143 & 0.238 & 0.143 & 0.190 \\ 
  0.75 & 0.333 & 0.333 & 0.333 & 0.286 & 0.286 & 0.238 & 0.238 & 0.190 & 0.143 \\ 
  1.00 & 0.143 & 0.143 & 0.095 & 0.143 & 0.095 & 0.095 & 0.095 & 0.095 & 0.095 \\ 
  3.00 & {\bf 0.048} & {\bf 0.048} & {\bf 0.048} & {\bf 0.048} & {\bf 0.048} & {\bf 0.048} & {\bf 0.048} & 0.095 & {\bf 0.048} \\ 
  5.00 & 0.095 & 0.095 & 0.095 & {\bf 0.048} & {\bf 0.048}& {\bf 0.048} &{\bf 0.048} & 0.095 & 0.095 \\ 
  \hline
\end{tabular}
\label{tab_mis}
\end{table}

\bibliographystyle{agsm} \bibliography{ref1}


\end{document}